\documentclass[submission, Phys]{SciPost}
\usepackage{hyperref}
\usepackage{amsmath}
\usepackage{amssymb}
\usepackage{graphicx}
\usepackage{color}
\usepackage{subfig}
\usepackage{bm}

\begin{document}

\begin{center}{\Large \textbf{
Relativistic Landau quantization in non-uniform magnetic field and its applications to
white dwarfs and quantum information
}}\end{center}

\begin{center}
Srishty Aggarwal\textsuperscript{1*},
Banibrata Mukhopadhyay\textsuperscript{1*},
Gianluca Gregori\textsuperscript{2*}
\end{center}

\begin{center}
{\bf 1} Department of Physics, Indian Institute of Science, Bangalore 560012, India

{\bf 2} Department of Physics, University of Oxford, Parks Road,
Oxford OX1 3PU, UK 
\\
* srishtya@iisc.ac.in,
* bm@iisc.ac.in,
* gianluca.gregori@physics.ox.ac.uk
\end{center}

\begin{center}
\today
\end{center}


\section*{Abstract}
{\bf
	We investigate the two-dimensional motion
 of relativistic cold electrons in the presence of `strictly' spatially varying magnetic fields
 satisfying, however, no magnetic monopole condition.
 We find that the degeneracy of Landau levels, which 
arises in the case of the constant magnetic field, lifts out when the field is
variable and the energy levels of spin-up and spin-down electrons align in an interesting way depending on the nature of change of field. Also the varying magnetic field splits Landau levels
	of electrons with zero angular momentum from positive 
	angular momentum, unlike the constant field which only can
	split the levels between positive and negative angular momenta.
	Exploring Landau quantization in non-uniform magnetic fields is
	a unique venture on its own and has interdisciplinary implications in the fields ranging 
from condensed matter to astrophysics to quantum information. As examples, we show 
	magnetized white dwarfs, with varying magnetic fields, involved simultaneously with Lorentz force and 
Landau quantization affecting the underlying degenerate electron gas, exhibiting a significant violation 
of the Chandrasekhar mass-limit; and an increase in quantum speed of electrons in the presence of a spatially growing magnetic field.

}

\vspace{10pt}
\noindent\rule{\textwidth}{1pt}
\tableofcontents\thispagestyle{fancy}
\noindent\rule{\textwidth}{1pt}
\vspace{10pt}

\section{Introduction}
The role of magnetic fields in controlling the natural -- Earth based to 
astrophysical -- systems from 
the microscopic to macroscopic scales is well established. 
From the formation of stars to stellar winds, cosmic rays, 
accretion disks and jets in X-ray binaries and active galactic nuclei, the magnetic field plays an indispensable role in all the astrophysical systems. In the Earth 
based systems and laboratory, quantum Hall effect, de Haas Van Alphen 
effect, vortices, superconductivity, high-resolution NMR and EPR spectroscopies
are some of the landmark contributions of high magnetic field physics to the
solid state and condensed matter sciences, analytical chemistry and structural biology.
 
The interaction of strong magnetic field with Fermi gas gives rise to many interesting effects. Two of the main effects are Landau Quantization (hereinafter LQ) \cite{1930ZPhy...64..629L} and Geometric Phase \cite{1984RSPSA.392...45B}. Most of the other applications appear to be advanced manifestations of these two. LQ has been well established and discussed in detail for uniform magnetic fields in both non-relativistic \cite{1965qume.book.....L} as well as relativistic \cite{1998rqm..book.....S} cases. In one hand, non-relativistic Landau effect has been extremely useful in explaining many condensed matter experiments through, e.g., quantum Hall effect, de Haas Van Alphen effect and Shubnikov-de Haas oscillations (see, e.g.,\cite{doi:10.1080/14786440908521019,1978JPSJ...44.1839W, 2001PhyB..299....6M, 2012PhRvL.108u6803C}).
On the other hand, relativistic LQ is helpful in resolving many astrophysical 
mysteries and quantum speed limit of fermions (e.g. \cite{PhysRevA.92.042106}). 
Effect of high magnetic field in neutron stars, particularly in
the surface with magnitude $\sim 10^{15}$ G as is proposed in the premise of
magnetar based model, is involved with LQ (even if involved with many levels).
Further, Das \& Mukhopadhyay by taking the
stoke of LQ explained the possible existence of super-Chandrasekhar mass white 
dwarfs \cite{PhysRevD.86.042001} 
and their new mass-limit \cite{2013PhRvL.110g1102D}, assuming magnetic fields to be uniform in 
such white dwarfs. It was also shown that LQ leads to softening the equation 
of state for neutron stars in the presence of strong magnetic field, though
the stiffening effect due to anomalous magnetic moment may overwhelm it
\cite{2000ApJ...537..351B}.
In addition, it was shown by one of the present authors that strong magnetic
field induced LQ influences the neutronization threshold and the onset of neutron 
drip by increasing the density for the former and increasing or decreasing for the latter
depending on the magnetic field \cite{2014PhRvC..89f5804V}. It
was further confirmed by others \cite{2015PhRvC..91f5801C} showing that the neutron drip line in the crust of highly magnetized star 
shifts to either higher or lower densities depending on the magnetic field 
strength.
Interestingly, synthetic Landau levels for photons has also been explored \cite{2016Natur.534..671S}. 
As photons experiencing a Lorentz force develop handedness, they provide opportunities
to study quantum Hall physics and topological quantum science.

However, most of the LQ effects mentioned above are probed 
maintaining the field uniform. In reality, particularly, in the astrophysical cases 
pointed out above, the magnetic field is never uniform. Note that LQ effects
become important only when the gyromagnetic
radius is comparable or less than the Compton
wavelength of the underlying particles. Moreover, LQ theory based on uniform magnetic field
does not suit for non-uniform magnetic fields, if
the magnetic field varies in a length scale comparable or shorter than the Compton wavelength of
the particles.

Altshuler \& Ioffe
\cite{PhysRevLett.69.2979} were the first to discuss the motion of fast particles in the strongly fluctuating magnetic field and showed analytically how 
fluctuations result in phase incoherence. Following their work, others
(see, e.g., \cite{PhysRevB.47.12051,PhysRevB.53.12917}) discussed multiple aspects of the motion 
of particles in magnetic fields, taking into considerations strong as 
well as weak but random fluctuations over uniform field, spatially modulated magnetic 
fields lifting degeneracy in the nonrelativistic regime, etc. 
There are other explorations of the effects of random magnetic fields 
to the electron gas and LQ, and their implications to, e.g., composite fermions 
and the quantum critical point \cite{PhysRevB.77.115353,PhysRevB.100.014442}.
 
What if, the field is completely non-uniform like in astrophysical systems and plasma?  
In white dwarfs, neutron stars, as well as main sequence stars, e.g. in Sun, 
it is almost certain that field varies from the center to surface. If the 
magnetar is a highly magnetized neutron star, while its surface field 
is observationally inferred to be $\sim 10^{15}$ G, its central field could be 
orders of magnitude higher. Similarly, highly magnetized white dwarfs (B-WDs)
violating Chandrasekhar-limit significantly with a new mass-limit 
\cite{2013PhRvL.110g1102D} are argued to have central and surface fields $\geq 10^{15}$ G and $\leq 10^{12}$ G respectively \cite{2015MNRAS.454..752S,2019MNRAS.490.2692K,2020ApJ...896...69K,
2020MNRAS.496..894G}, hence LQ clearly can not be avoided therein.
Note that at high densities in B-WDs, the Coulomb interactions turn out to be negligible compared to Fermi energy \cite{PhysRevD.86.042001}. Therefore electric fields are negligible compared to the magnetic fields therein, particularly if the field is not time varying. Hence, one can ignore the QED induced effect of pair creation in
B-WDs \cite{1968ApJ...153L.157C,PhysRevD.14.340,PhysRevD.86.042001}, which
can not happen in magnetic fields alone.
Indeed, field magnitude decaying with density was proposed earlier for 
neutron stars and white dwarfs \cite{PhysRevLett.79.2176,2014JCAP...06..050D} in an
ad-hoc basis and predicted its influence on the mass, radius, and luminosity
(see, e.g., \cite{2020MNRAS.496..894G}, for a latest application). However, there was no 
account for the associated potential and Maxwell's equations in such variation 
of magnetic field. Hence the effect of magnetic field was considered in an
adiabatic approximation and hence LQ in uniform field.
However, such systems with 
magnetic field having variation over spatial region as an explicit 
function of distance has not gained direct attention till date. 

We aim here to 
explore the change in LQ effect on the energy levels of relativistic 
electrons in the presence of mainly a decaying magnetic field, but
also a growing magnetic field. The chosen field 
profiles, as demonstrated below, are in accordance with Maxwell's equations 
of electromagnetism. Since LQ is a quantum phenomenon, it is expected to 
be affected only if the variation in magnetic field takes place at the quantum 
scale. Therefore, we probe the variation of magnetic field at scales 
of the order of  gyromagnetic radius which is less than or of the order of Compton wavelength of electron, determined by the chosen magnetic fields.

It is generally expected that the magnetic field and density vary in a star as decreasing functions of its radial coordinate. Therefore, the
magnetic field is expected to be varying with the density. As the density is expected to be highest at the center and lowest at the 
surface, it is a reasonable assumption that the magnetic field too follows the same trend, as proposed earlier \cite{PhysRevLett.79.2176}. 
Such a field profile has been extensively used for neutron stars and white dwarfs with appropriate parameters (see, e.g., \cite{10.1093/mnras/sty776,2020MNRAS.496..894G}), which may induce a sharp variation of the field in a short spatial scale, depending upon the parameters.

As an application of non-uniform and growing magnetic field, we also show its role in attaining higher quantum speed, i.e. transition speed from one energy level to other, 
of electrons as compared to uniform magnetic field. This could help in achieving faster processing speed of quantum computers along with other applications.

For relativistic electrons, the splitting 
of levels due to spin has significant contribution in determining the energy 
spacing and overall structure of energy levels and, hence, cannot be treated 
perturbatively. For uniform magnetic field, it, in fact, leads to doubly 
degenerate levels \cite{1965qume.book.....L}. We show and investigate, how the degeneracy due to 
spin arising in constant magnetic field breaks down when the field is variable.

In the next section (Sec.\ref{sec2}), we 
establish the formalism of the problem of variable magnetic field.
Subsequently, we review 
the solution for uniform magnetic field in Sec.\ref{sec3} and explore the effect
of the non-uniform magnetic field in detail including the effective potential 
observed by the electron in Sec.\ref{sec4}. The computational 
methods to determine the eigenvalue spectrum are also outlined in Sec.\ref{sec4}. 
The solutions of established equations are shown in Sec.\ref{sec5} and 
Sec.\ref{sec6}. The underlying thermodynamics and equation of state (EoS) are
explored in Sec.\ref{eos} and its implications, in an 
astrophysical context and a quantum information, have been enlightened in Sec.\ref{appl}.
We conclude in Sec.\ref{sec7} by highlighting the key points 
of this work and its various implications.

\section{\label{sec2} Dirac equation for electrons and its solution in the presence of magnetic fields}
For electron of mass $m_{e}$ and charge $q~(-e)$, the Dirac equation in the 
presence of magnetic field is given by
\begin{equation}
i\hbar\frac{\partial\Psi}{\partial t} = \left[ c\boldsymbol{\alpha}\cdot\left(-i\hbar\textbf{$\nabla$}-\frac{q\textbf{A}}{c}\right) + \beta m_{e}c^2\right]\Psi,
\label{eq1}
\end{equation}
where $\boldsymbol{\alpha}$ and $\beta$ are Dirac matrices, $\textbf{A}$ 
is the vector potential, $\hbar=h/2\pi$ with $h$ being Planck's constant and 
$c$ is the speed of light. For stationary states, we can write
\begin{equation}
	\Psi = e^{-i\frac{Et}{\hbar}}\begin{bmatrix}
  \chi \\
 \phi \\
 \end{bmatrix}
\label{matrix},
 \end{equation} 
where $\Phi$ and $\chi$ are 2-component objects/spinors. We consider the Pauli-Dirac representation in which
\begin{equation}
\alpha = \begin{bmatrix}
0 & \boldsymbol\sigma\\
\boldsymbol\sigma & 0\\
\end{bmatrix}
~,\beta = \begin{bmatrix}
I & 0\\
0 & -I\\
\end{bmatrix},
\end{equation}
where each block represents a $2\times 2$ matrix and $\boldsymbol{\sigma}$ 
represents three components of the Pauli matrices together in a vector.
Hence Eq. (\ref{eq1}) reduces to
\begin{equation}
(E-m_{e}c^2)\chi = c\textbf{$\sigma$}\cdot\left(-i\hbar\textbf{$\nabla$}-\frac{q\textbf{A}}{c}\right)\phi,
\end{equation}
\begin{equation}
(E+m_{e}c^2)\phi = c\boldsymbol\sigma\cdot\left(-i\hbar\textbf{$\nabla$}-\frac{q\textbf{A}}{c}\right)\chi.
\end{equation}
Decoupling them for $\chi$, we obtain
\begin{equation}
(E^2-m_{e}^2c^4)\chi =\left[c\boldsymbol\sigma\cdot\left(-i\hbar\textbf{$\nabla$}-\frac{q\textbf{A}}{c}\right)\right]^2\chi.
\label{eq2}
\end{equation}
Defining $\boldsymbol{\pi}=-i\hbar\textbf{$\nabla$}-q\textbf{A}/c$ and using the identity $(\boldsymbol\sigma\cdot\boldsymbol\pi)(\boldsymbol\sigma\cdot\boldsymbol\pi) = \pi ^2 - q\hbar\boldsymbol\sigma\cdot \boldsymbol{B}/c$, Eq. (\ref{eq2}) reduces to
\begin{equation}
(E^2-m_{e}^2c^4)\chi = \left[c^2\left(\pi ^2 - \frac{q\hbar}{c}\boldsymbol\sigma\cdot \textbf{B}\right)\right]\chi, 
\label{eq3}
\end{equation}
such that the antiparticle wavefunction $\phi=-\chi$ \cite{1998rqm..book.....S}. We solve Eq. (\ref{eq3}) for a variable magnetic field in cylindrical coordinates. As there is no fixed law for the variation of magnetic field in nature,
except that it should satisfy Maxwell's equations, we choose a simple 
power law variation of the magnetic field, given by
\begin{equation}
\textbf{B} = B_{0}\rho^n \hat{z},
\end{equation}
in cylindrical coordinates $(\rho,\phi,z)$. Such a field profile satisfies no monopole condition 
\textbf{$(\nabla\cdot\boldsymbol{B} = 0)$} and according to Amp\'ere's law
produces current. See appendix for total Lagrangian and from Lagrangian
equation of motion how to obtain the Dirac and Maxwell's equations. For
the present purpose of underlying quantum physics, our interest is in 
the Dirac equation. However, in certain applications, e.g. in stellar physics,
the underlying Maxwell's equation needs to be paid attention in order to include classical Lorentz force. The chosen
magnetic field profile also assures the 
decaying nature of the field away from the source if $n<0$ which is a 
common feature, particularly in stellar physics. Also, the same profile with $n>0$ can be applicable for
a system with spatially growing field satisfying other physics intact.
Using a gauge freedom for the vector potential \textbf{A}, we choose
\begin{equation}
 \textbf{A} = B_{0}\frac{\rho^{n+1}}{n+2} \hat{\phi} = A\hat{\phi}.
\end{equation}
Hence,
\begin{equation}
  \pi ^2\chi=\left[\hat{p}_{\rho}^2+\left(\hat{p}_{\phi} - \frac{qA}{c}\right)^2+\hat{p}_{z}^2\right]\chi,
\label{eq4}
\end{equation} 
where $\hat{p}_{\rho,\phi,z}$ denote operators.
Noticing that $\phi$ and $z$ are ignorable coordinates, the solution of Eq. (\ref{eq3}) can be written as
\begin{equation}
 \chi = e^{i\left(m\phi+\frac{p_{z}}{\hbar}z\right)}R(\rho),
 \end{equation}
where $R(\rho)$ is a two-component matrix, `$m\hbar$' is the angular momentum 
of the electron and $p_{z}$ is the eigenvalue of momentum in the $z-$direction. Therefore, Eq. (\ref{eq4}) becomes
\begin{equation}
\pi ^2 R = -\hbar ^2 \left[\frac{\partial ^2}{\partial \rho^2}+\frac{1}{\rho}\frac{\partial}{\partial\rho}-\frac{m^2}{\rho^2}\right]R(\rho)
+\left[\frac{q^2A^2}{c^2}+\frac{2q\hbar mA}{c\rho}+p_{z}^2\right]R(\rho).
\label{eq5}   
\end{equation} 
From Eqs. (\ref{eq3}), (\ref{eq4}) and (\ref{eq5}) and substituting $q = -e$, we obtain
\begin{multline}
\left(\frac{E^2 - m_{e}^2c^4}{c^2}-p_{z}^2\right)R(\rho) =  -\hbar ^2 \left[\frac{\partial ^2}{\partial \rho^2}+\frac{1}{\rho}\frac{\partial}{\partial\rho}-\frac{m^2}{\rho^2}\right]R(\rho)\\
+\left[\frac{e^2A^2}{c^2}-\frac{2e\hbar mA}{c\rho}+\frac{e\hbar}{c}(\sigma_{z}B)\right]R(\rho).
\label{eq6}
\end{multline}
There will be two independent solutions for $R(\rho)$, which can be taken, without loss of generality, to be the eigenstates of $\sigma_{z}$, with eigenvalues $\pm 1$. Thus if we choose two independent solutions of the form
\begin{equation}
 R_{+}(\rho)=\begin{bmatrix}
 \tilde{R}_{+}(\rho)\\
 0
 \end{bmatrix}
~,R_{-}(\rho)=\begin{bmatrix}
 0\\
 \tilde{R}_{-}(\rho)\\
 \end{bmatrix}
 \nonumber
 \end{equation}
such that $\sigma_{z}{R}_{\pm}=\pm{R}_{\pm}$, Eq. (\ref{eq6}) becomes

\begin{equation}
\tilde{P}\tilde{R}_{\pm}= -\hbar ^2 \left[\frac{\partial ^2}{\partial \rho^2}+\frac{1}{\rho}\frac{\partial}{\partial\rho}-\frac{m^2}{\rho^2}\right]\tilde{R}_{\pm}+\left[\frac{e^2A^2}{c^2}-\frac{2e\hbar mA}{c\rho}\pm\frac{e\hbar}{c}B\right]\tilde{R}_{\pm} \label{eq8}
\end{equation}
where
\begin{equation}
\tilde{P} = \left(\frac{E^2 - m_{e}^2c^4}{c^2}-p_{z}^2\right).
\label{eqP}
\end{equation}
Dividing Eq. (\ref{eq8}) by $m_{e}^2c^2$, we have an eigenvalue equation as
\begin{align}
\alpha\tilde{R}_{\pm} &= -\left(\frac{\hbar}{m_{e}c}\right)^2 \left[\frac{\partial ^2}{\partial \rho^2}+\frac{1}{\rho}\frac{\partial}{\partial\rho}-\frac{m^2}{\rho^2}\right]\tilde{R}_{\pm} 
+\left[\frac{e^2A^2}{m_{e}^2c^4}+\frac{e\hbar}{m_{e}^2c^3}\left(-\frac{2mA}{\rho}\pm B\right)\right]\tilde{R}_{\pm} \label{eq9}\\ 
  &= -\lambda_{e}^2\left[\frac{\partial ^2}{\partial \rho^2}+\frac{1}{\rho}\frac{\partial}{\partial\rho}-\frac{m^2}{\rho^2}\right]\tilde{R}_{\pm}  + \left[\left(\frac{kB_0\rho^{n+1}}{n+2}\right)^2 + k\lambda_{e}\left(-\frac{2m}{n+2}\pm 1\right)B_0\rho^n\right]\tilde{R}_{\pm},
\label{eq10}
\end{align}
where $\alpha = \frac{\tilde{P}}{m_{e}^2 c^2} = (\epsilon^2 - 1-x_{z}^2)$
which is, in fact, square of dimensionless momentum and acting as an eigenvalue of the problem,
$\epsilon = \frac{E}{m_{e}c^2}$ (dimensionless energy),
$x_{z} = \frac{p_{z}}{m_{e}c}$\,\; (dimensionless momentum along $z-$direction),
$\lambda_{e} = \frac{\hbar}{m_{e}c}$ (Compton wavelength of electrons),
$k=\frac{e}{m_{e}c^2}$.
Note that this $\alpha$ should not be confused with Dirac $\boldsymbol{\alpha}$ matrix.

\subsection{\label{sec3}Uniform Magnetic Field ($n = 0$)}

For constant magnetic field, Eq. (\ref{eq10}) becomes
\begin{equation}
\alpha\tilde{R}_{\pm} = -\lambda_{e}^2\left[\frac{\partial ^2}{\partial \rho^2}+\frac{1}{\rho}\frac{\partial}{\partial\rho}-\frac{m^2}{\rho^2}\right]\tilde{R}_{\pm} + \left[\left(\frac{kB_0\rho}{2}\right)^2 + k\lambda_{e}\left(-m\pm 1\right)B_0\right]\tilde{R}_{\pm}.
\label{eq11}
\end{equation}
The above equation can be solved analytically similar to its non-relativistic counterpart \cite{1965qume.book.....L}.
Now defining $\xi = \left(\frac{kB_{0}}{2\lambda_{e}}\right)\rho^2$, Eq. (\ref{eq11}) can be written as
\begin{equation}
\xi \tilde{R}_\pm'' + \tilde{R}_\pm'+\left(-\frac{1}{4}\xi + \beta_{\mp} - \frac{m^2}{4\xi}\right)\tilde{R}_\pm = 0,
\label{xieq}
\end{equation}
where
\begin{equation}
\beta_{\mp} = \frac{\alpha}{2\lambda_{e}kB_{0}}+\left(\frac{m}{2}\mp\frac{1}{2}\right)
\nonumber
\end{equation}
and double-prime ($''$) and prime ($'$) respectively denote double and single
derivatives with respect to $\rho$.
At $\xi\rightarrow\infty$, the solution of Eq. (\ref{xieq}) gives as
$\tilde{R}_\pm\sim e^{-\frac{\xi}{2}}$, and for $\xi\rightarrow0$ as 
$\tilde{R}_\pm\sim\xi ^{\frac{|m|}{2}}$. Accordingly, we seek a solution of 
the form
\begin{equation}
\tilde{R}_\pm = e^{-\frac{\xi}{2}}\xi ^{\frac{|m|}{2}} w(\xi).
\end{equation}
Thence equation for $w(\xi)$ satisfies the confluent hypergeometric function so that
\begin{equation}
w = F\left[-\left(\beta_{\mp} -\frac{|m|}{2}-\frac{1}{2}\right),|m|+1,\xi\right].
\end{equation}
For the wavefunction to be finite everywhere, the quantity $\left(\beta_{\mp} -\frac{|m|}{2}-\frac{1}{2}\right)$ must be a non-negative integer $\nu$. Hence, the values of $\alpha$ are given by
\begin{equation}
\alpha_{\nu} = 2k\lambda_{e}B_{0}\left (\nu+\frac{|m|}{2}-\frac{m}{2}+\frac{1}{2}\pm\frac{1}{2}\right),
\label{eq12}
\end{equation}
where $m$ is the azimuthal quantum number.
One can easily see from Eq. (\ref{eq12}) that ground state energy (corresponding to
$\alpha_{0}$) is 0 and all the other energy levels are doubly degenerate. 
Also, energies are same for $m=0$ and $>0$.
Finally $\tilde{P}$ in Eq. (\ref{eqP}) turns out to be $2\nu B_0/B_c$, where
$B_c=m_e^2c^3/e\hbar$, the Schwinger limit of pair production, so that
\begin{equation}
E^2=p_z^2c^2+m_e^2c^4\left(1+2\nu\frac{B_0}{B_c}\right).
\label{Ev}
\end{equation}

\subsection{\label{sec4}Non-Uniform Magnetic Field ($n\neq0$)} 
We know that analytic solutions exist for some special potentials only, which include harmonic oscillator, hydrogen-atom and Morse-oscillator.
For the presently chosen potential, however, we are not able to find solutions analytically. Therefore, we use computational methods to find eigenvalues $\alpha_\nu$ at different levels $\nu$ for different $n$. Let us first explore the asymptotic behaviour of $\tilde{R}_\pm$ (the asymptotic behaviour is same for  $\tilde{R}_+$ and  $\tilde{R}_-$).

As $\rho\rightarrow 0$, Eq. (\ref{eq10}) becomes
\begin{equation}
-\lambda_e^2\left[\frac{\partial^2}{\partial\rho^2}+\frac{1}{\rho}\frac{\partial}{\partial\rho}\right]\tilde{R}_\pm=0.
\end{equation}
Hence, $\tilde{R}_\pm\rightarrow C_1+C_2\log(\rho)$, $C_1$ and $C_2$ being constants. Since $\log(\rho)$ blows up at $\rho\rightarrow 0$, to seek for a finite 
solution throughout, we set $C_2=0$. Hence, as $\rho\rightarrow 0$
\begin{equation}
\tilde{R}_\pm\rightarrow C_1,\quad \tilde{R}_\pm'\rightarrow 0.
\label{eq13}
\end{equation}  

For $\rho\rightarrow\infty$, however, for $n\leq0$, Eq. (\ref{eq10}) turns out to be
\begin{equation}
	\left[-\lambda_e^2\frac{\partial^2}{\partial\rho^2}+\left(\frac{kB_0\rho^{n+1}}{n+2}\right)^2\right]\tilde{R}_\pm=0.
\end{equation}
Thus,
\begin{equation}
\tilde{R}_\pm \rightarrow e^{-\left[\frac{kB_0}{\lambda_e(n+2)}\right]\frac{\rho^{n+2}}{n+2}}\quad {\rm as}\:\;\rho\rightarrow \infty.
\end{equation}

\begin{figure}[hbtp]
\includegraphics[scale=0.7]{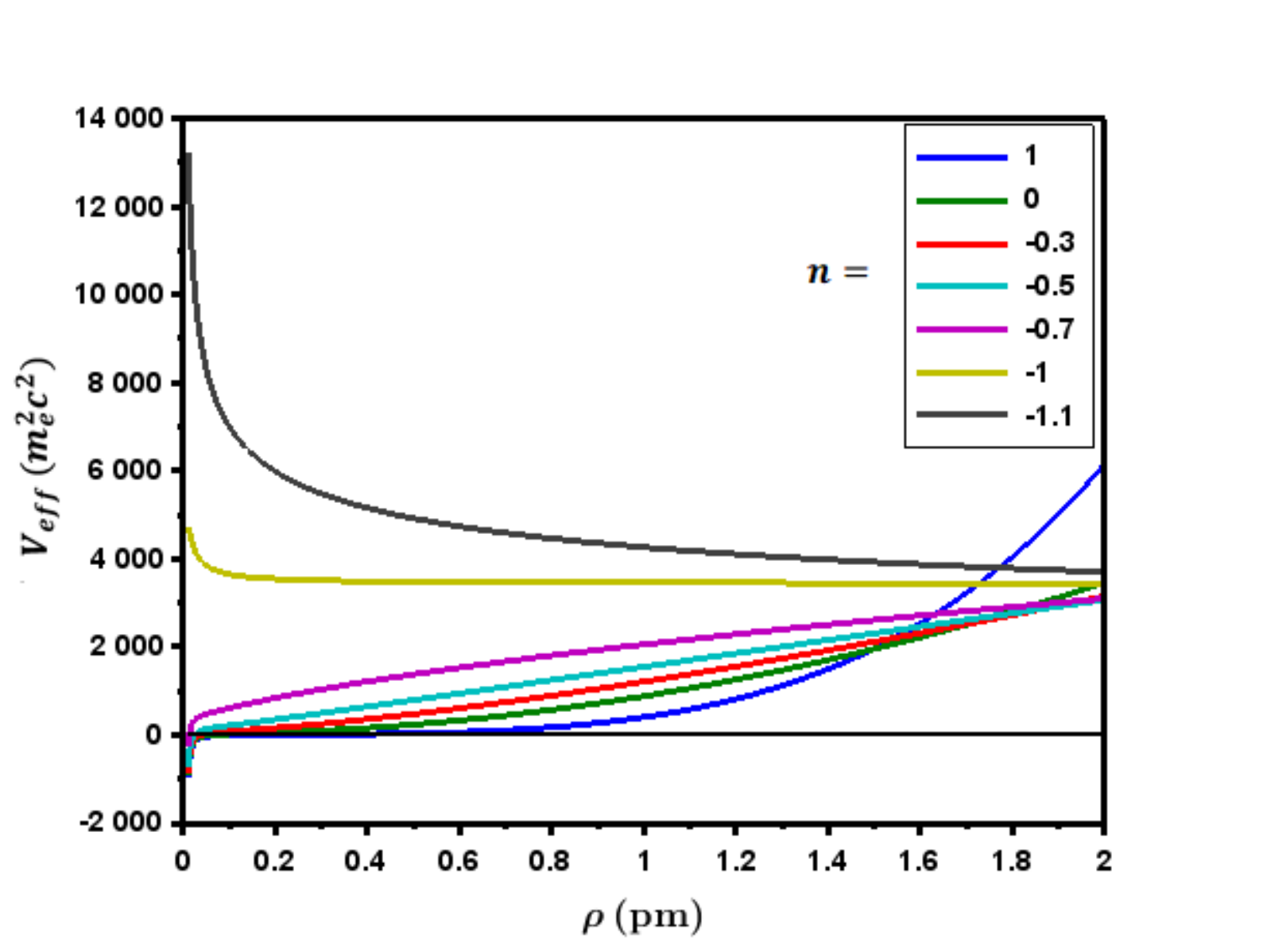}
\centering
\caption{The variation of effective potential for different $n$ for
$B_0 = 10^{15}$ G pm$^{-n}$. Here, the black horizontal line represents 
$V_{eff}=0$. Various potentials at $\rho=1$~pm from bottom to top successively
	are for $n=1,0,-0.3,-0.5,-0.7,-1,-1.1$.}
\label{veff}
\end{figure}

There are many different methods to solve Eq. (\ref{eq10}) including 
`Finite Difference' method and `Shooting and Matching' method. We obtain most 
accurate solutions with the `Shooting and Matching' method \cite{2002nrca.book.....P}, where the relative
error between results in exact theory and computation is below 0.0004
for lower Landau levels and it never exceeds 0.002 even in higher levels 
for the constant field case (see Table \ref{table1}). The differential equation is solved using ``ode rk" command in scilab 
\footnote{see, https://help.scilab.org/docs/6.0.0/en$\_$US/ode.html}
 which is traditional adaptive Ranga-Kutta method. In order to 
obtain the initial conditions, we use our knowledge for the behavior of 
$\tilde{R}_\pm$ at $\rho\rightarrow 0$ and then set $C_1 = 1$. Thus, we have initial 
conditions as $\tilde{R}_\pm(\rho\rightarrow 0)=1$, $\tilde{R}_\pm' (\rho\rightarrow 0)=0$. Ideally, initial conditions should be defined at $\rho=0$, but 
many terms in Eq. (\ref{eq11}) blows up at $\rho=0$. Hence, we define the 
initial conditions at $\rho= 10^{-10}$ picometer (pm)  which is equivalent to 0 compared to even
the minimum gyromagnetic radius for our field of
interest, which is of the order of pm. We express magnetic fields in units of G and length in pm for solving Eq. (\ref{eq10}). 
Thus, $B_{0} = |\boldsymbol{B}|=B$ at 1 pm.

Also, to remove the diverging nature of magnetic field near the origin 
with $n<0$, we choose 
\begin{equation}
B = B_0\:(\rho+\rho_0)^n,
\end{equation}
where $\rho_0$ could be chosen to be a very small number as compared to the scale of 
wavefunction. We choose it to be equal to $10^{-5}$ pm. As long as $\rho_0$ is 
very small, this choice does not effect the solutions.

To determine the effective potential experienced by electrons, let $\tilde{R}_\pm(\rho) = \frac{u_\pm(\rho)}{\sqrt{\rho}}$. Then, Eq. (\ref{eq10}) becomes
\begin{equation}
\alpha u_\pm = \left(-\lambda_{e}^2\frac{\partial ^2}{\partial \rho^2}+V_{eff}\right)u_\pm
\label{eq20}
\end{equation}
where
\begin{equation}
V_{eff} = -\lambda_{e}^2\left[\frac{1}{4\rho^2}-\frac{m^2}{\rho^2}\right]
\nonumber + \left(\frac{kB_0\rho^{n+1}}{n+2}\right)^2 + k\lambda_{e}\left(-\frac{2m}{n+2}\pm 1\right)B_0\rho^n.
\label{eq21}
\end{equation}

We show the variation of $V_{eff}$ for different $n$ in Figure \ref{veff}. 
It is seen that for $n\leq-1$, potential is completely repulsive whose 
solution will depend on the distance from the source (origin of the system) 
upto which a particle 
can move. Therefore, the energy eigenvalues for such cases depend upon 
where we put a hard wall making the system equivalent to confining 
the electron in a box. However we do not want to apply any such restrictions 
on the electron. Moreover, this nature of variation is not realistic, particularly in astrophysical scenarios. We therefore restrict our analysis to cases for 
$n>-1$.

\begin{table}
\centering

\begin{tabular}{|c|l|l|l|}
\hline
$\:\;\nu\:\;$ & $\alpha _{comp}$ & $\alpha_{th}$ & relative error\\
\hline
0 & 22.200623$\quad$ & 22.2094$\quad$ &   0.0003952\\
1 & 66.616364 & 66.6282 & 0.0001776\\
2 & 111.03531 & 111.047 & 0.0001053\\
3 & 155.4541 & 155.4658 & 0.0000752\\
4 & 199.87289 & 199.8846 & 0.0000586\\
5 & 244.29169 & 244.3034 & 0.0000479\\
6 & 288.7104 & 288.7222 & 0.0000409\\
7 & 333.12916 & 333.141 & 0.0000355\\
8 & 377.54795 & 377.5598 & 0.0000314\\
9 & 421.96673 & 421.9786 & 0.0000281\\
\hline
\end{tabular}

\caption{Comparison of the eigenvalues obtained from numerical computation
	(column two) and theory (column three) for constant magnetic fields ($n=0$) with $B_0 = 10^{15}\:G~{\rm pm}^{-n}$.}
\label{table1}
\end{table}

\section{Dispersion relations}
\subsection{\label{sec5}Excluding Zeeman effect}
\begin{figure}
\centering
\includegraphics[scale=0.7]{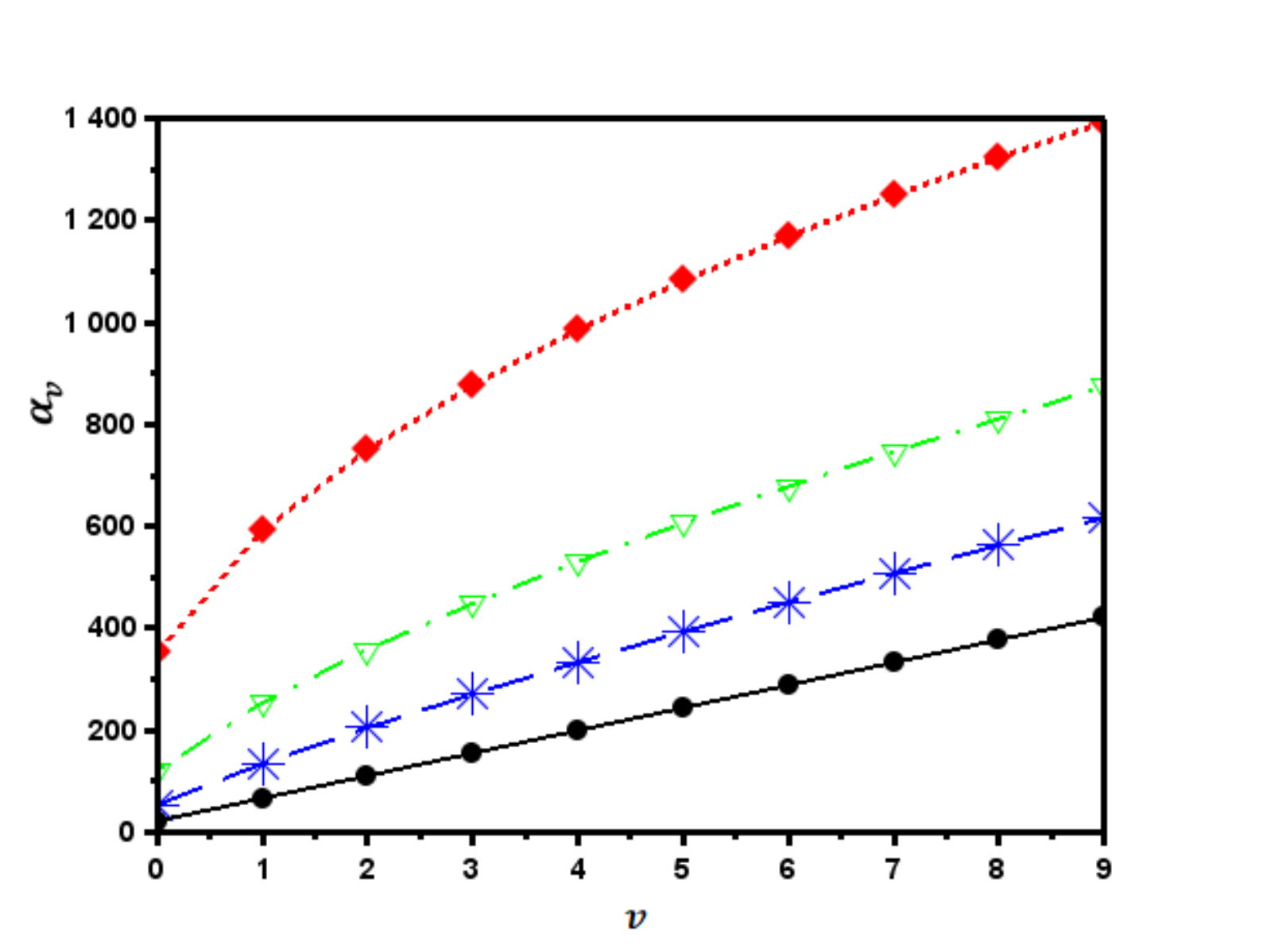}
\caption{The variation of eigenvalue with eigen-index for $B_0 = 10^{15}$ 
G pm$^{-n}$, when $n = 0$ (black solid circles), -0.3 (blue dashed asterisks), -0.5 (green dot-dashed triangles) and -0.7 (red dotted diamonds). The lines represent the curves fitted with constants of Eq. (\ref{eq14}).}
\label{fig without}
\end{figure}

First, we investigate the effect of variation of magnetic fields on the 
energy levels $\alpha_\nu$ excluding Zeeman splitting for $m=0$. Thus, Eq. (\ref{eq10}) becomes
\begin{equation}
\alpha'_\nu \tilde{R} = -\lambda_{e}^2\left[\frac{\partial ^2}{\partial \rho^2}+\frac{1}{\rho}\frac{\partial}{\partial\rho}\right]\tilde{R}
 + \left(\frac{kB_0\rho^{n+1}}{n+2}\right)^2 \tilde{R},
\end{equation}
where $\alpha'_\nu$ is the energy level excluding Zeeman effect. Figure \ref{fig without} shows how the spacing of energy levels modifies for different $n: -1< n\leq0$. 
As seen in the figure, as $n$ decreases, energy levels rise up. It is seen
from Figure \ref{veff} that with the decrease of $n$ (but for $>-1$), potential with increasing $\rho$ crosses the 0 
at a smaller distance and thereby becoming repulsive closer to the origin,
compared to that of a larger $n$. 
We know that a particle is more stable if it is in an attractive potential 
regime and has lower energy. If a particle feels a repulsive potential, 
it requires more energy to stay in that region, thus explaining the 
behavior of eigenvalues seen in Figure \ref{fig without}. In simpler words, 
this increase in energy eigenvalues can be understood as follows. These
variations of eigenvalues are for a fixed $B_0$, which is $B$ at 1~pm, when
$B$ keeps increasing to a much higher value near the source for lower $n$, 
thereby, increasing the average magnetic field and, 
hence, raising the energy levels.    

Also, with the decrease of $n$, dispersion energies become highly non-linear, 
i.e. the difference between two successive levels, while initially is very 
large, then decreases much faster for smaller $n$, which is physically 
related to the chosen profile of magnetic field. Due to the faster 
decaying nature of field, electrons observe a very strong magnetic field near 
the center, thereby, having significant discretion of energies, for a fixed
$B_0$. As it 
moves little away from the center, magnetic field weakens and, hence, the 
spacing of levels decreases. One can expect a larger change in the energy level 
gaps, if field decays more rapidly, what is seen in Figure \ref{fig without}.

\begin{table}
\centering

\begin{tabular}{|c|c|c|c|c|}
\hline
$n$ & $C_3$ & $C_4$ & $C_5$ & $C_6$\\
\hline
0 & $\:\;$44.4188$\:\;$ & 1 & 0.5 & 1\\
$\:\;$-0.1$\:\;$ & 56 & $\:\;$1.0519$\:\;$ & 0.50 & 0.9467\\
-0.2 & 72.5 & 1.111 &$\:\;$ 0.4934$\:\;$ & $\:\;$0.8878$\:\;$\\
-0.3 & 97 & 1.18 & 0.488 & 0.8224\\
-0.4 & 134.63 & 1.25 & 0.486 & 0.749\\
-0.5 & 195.66 & 1.33 & 0.484 & 0.665\\
-0.6 & 301 & 1.43 & 0.482 & 0.5702\\
-0.7 & 494  & 1.54 & 0.476 & 0.4609\\
-0.8 & 878.9 & 1.667 & 0.475 & 0.33\\
-0.9 & 1706 & 1.818 & 0.48 & 0.191\\
\hline
\end{tabular}

 \caption{The values of the constants of Eq. (\ref{eq14}) for various $n$. 
	Here $B_0$ in Eq. (\ref{eq14}) is chosen in the units of $10^{15} \:$ G pm$^{-n}$ to obtain $C_2$.}
\label{table2}
\end{table}

Since an analytical solution is not easy to obtain, we try to figure out the 
possible expression for the energy dispersion relation using a suitable ansatz 
and data fitting. Based on the analogy of constant field case, we suggest the 
ansatz of the form  
\begin{equation}
\alpha'_\nu= C_3\: B_0^{C_4}\:(\nu+C_5)^{C_6},
\label{eq14}
 \end{equation} 
 where $C_3,\:C_4,\:C_5$ and $C_6$ are constants whose values depend on $n$. 
Table ~\ref{table2} shows the values of these constants that we obtain by 
fitting numerical data for lower levels when $-0.9\le n\le 0$. 
It is interesting to note that
\begin{equation}
  C_4+C_6=2
\end{equation}  
and
\begin{equation}
    C_4(n) = \frac{2}{n+2}.
\end{equation}

\subsection{\label{sec6}Including Zeeman Effect}
\begin{figure}
\centering
\includegraphics[scale=0.45]{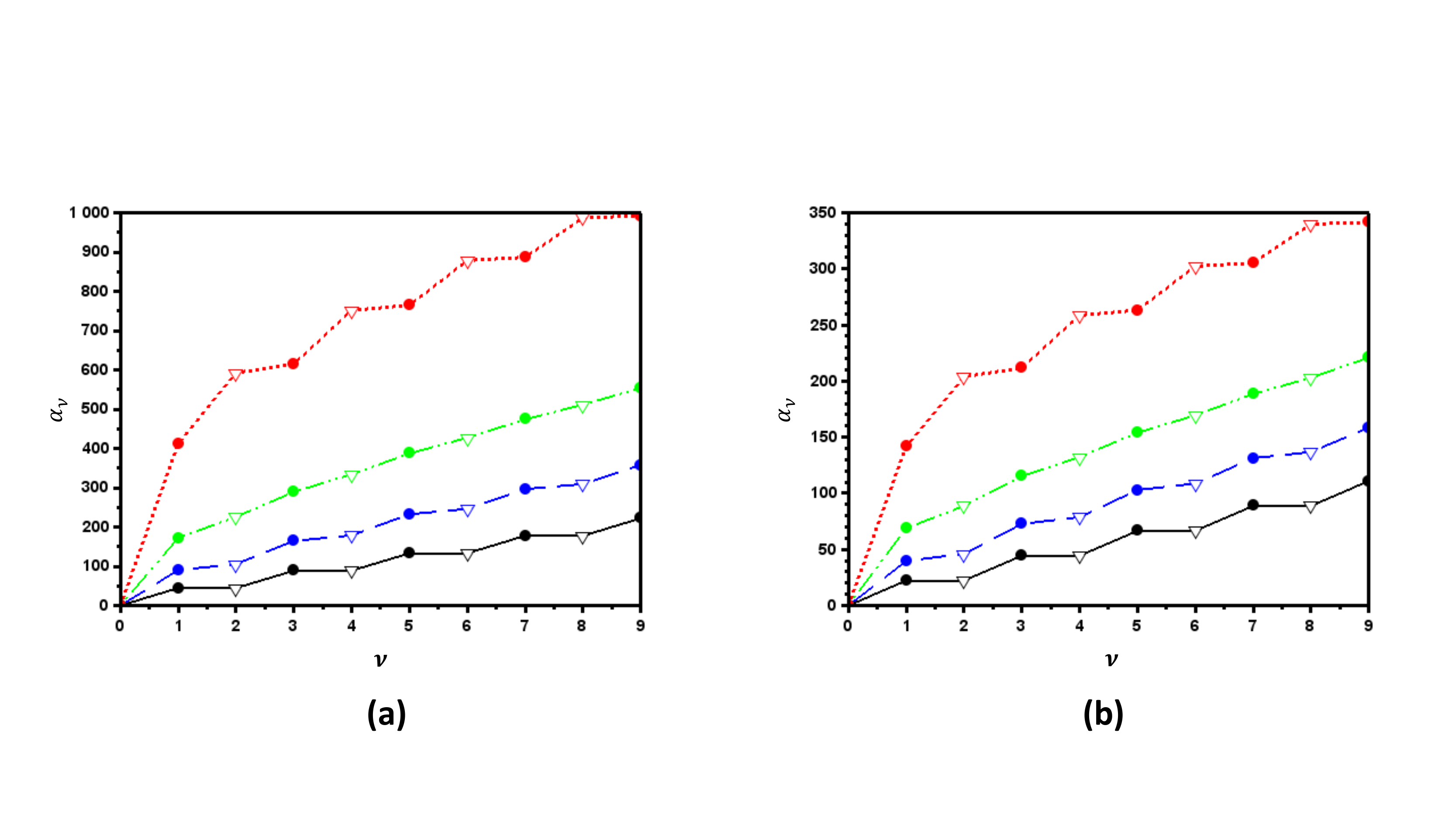}
\caption{The variation of eigenvalue with the eigen-index for $n =  0$ (black solid line), -0.3 (blue dashed line), -0.5 (green dot-dashed line) and -0.7 (red dotted line) for (a) $B_0 = 10^{15}$ G~pm$^{-n}$, and (b) $B_0 = 5\times 10^{14}$ G~pm$^{-n}$, and $m=0$. Here the levels for $-\boldsymbol{\sigma}\cdot\boldsymbol{B}$ ($-B_0$) and $+\boldsymbol{\sigma}\cdot\boldsymbol{B}$ ($+B_0$) are marked by the solid circles and triangles respectively.}
\label{figwith}
\end{figure}

\begin{table*}

\begin{tabular}{|c|c|c|cccc|}
\hline
	$n$ & $\quad \nu _m \quad$ & $B_0$\:($10^{15}$\:G~pm$^{-n}$) & $\alpha_0$ & $\alpha_1$
	& $\alpha_2$ & $\alpha_3$\\
\hline
0 & 1 & 8.98 & 0.00976 & 398.866 & 797.397 & 1196.62\\
  & 2 & 4.49 & 0.0152 & 199.461 & 398.879 & 598.311\\
  & 3 & 2.994 & 0.00976 & 133.0327 & 266.0217 & 398.98\\
\hline 
-0.3 & 1 & 3.546 & 0.0468 & 399.237 & (+)468.26 & 729.245\\
  & 2 & 3.095 & 0.00312 & 340.2 & (+)399.06 & 621.398\\
  & 3 & 2.125 & 0.002 & 218.28 & (+)256.37 & 399.255\\
\hline
-0.5 & 1 & 1.876 & 0.0428 & 399.15 & (+)532.63 & 669.2\\
  & 2 & 1.533 & 0.006 & 304.718 & (+)399.19 & 511.24\\
  & 3 & 1.275 & 0.0498 & 238.348 & (+)312.29 & 399.897\\
\hline
-0.7 & 1 & 0.979 & 0.000 & 399.44 & (+)573.177 & 595.61\\
  & 2 & 0.744 & 0.00217 & 278.27 & (+)399.3 & 413.934\\
  & 3 & 0.755 & 0.0003 & 267.83 & (+)384.34 & 399.345\\
 \hline     
\end{tabular}

\caption{The variation of $B_0$ for different $n$ for one-level, two-level and three-level systems at $\epsilon = 20$. For $n=0$, all the levels are doubly degenerate. For $n\neq0$ all eigenvalues are different independent of energy
levels and splits $+\boldsymbol\sigma\cdot\boldsymbol B$ and $-\boldsymbol\sigma.\boldsymbol B$. Here `(+)' denotes $\alpha_\nu$ with 
$+\boldsymbol\sigma\cdot\boldsymbol B$ and the rest is for $-\boldsymbol\sigma\cdot\boldsymbol B$.}
\label{table3}
\end{table*}
Now let us obtain the eigenvalues for the entire Eq. (\ref{eq10}). Figure 
\ref{figwith} shows the eigenvalues for $m=0$ with (a) $B_0 = 10^{15}$ G~pm$^{-n}$, and (b) $B_0 = 5\times 10^{14}$ G~pm$^{-n}$, where different markers distinctly
indicate the levels for $-\boldsymbol{\sigma}\cdot\boldsymbol{B}$ ($-B_0$)
and $+\boldsymbol{\sigma}\cdot\boldsymbol{B}$ ($+B_0$). To give a better idea 
about the variation of eigenvalues and the splitting of levels, we fix 
$\epsilon$ to the Fermi energy $\epsilon_F=20$ and then obtain $B_{0}$ and corresponding eigenvalues for one-level, 
two-level and three-level systems, enlisted in Table ~\ref{table3}.
There are many interesting results what can be inferred from Figure 
\ref{figwith} and Table ~\ref{table3}.

\begin{figure*}
\centering
\includegraphics[scale=0.7]{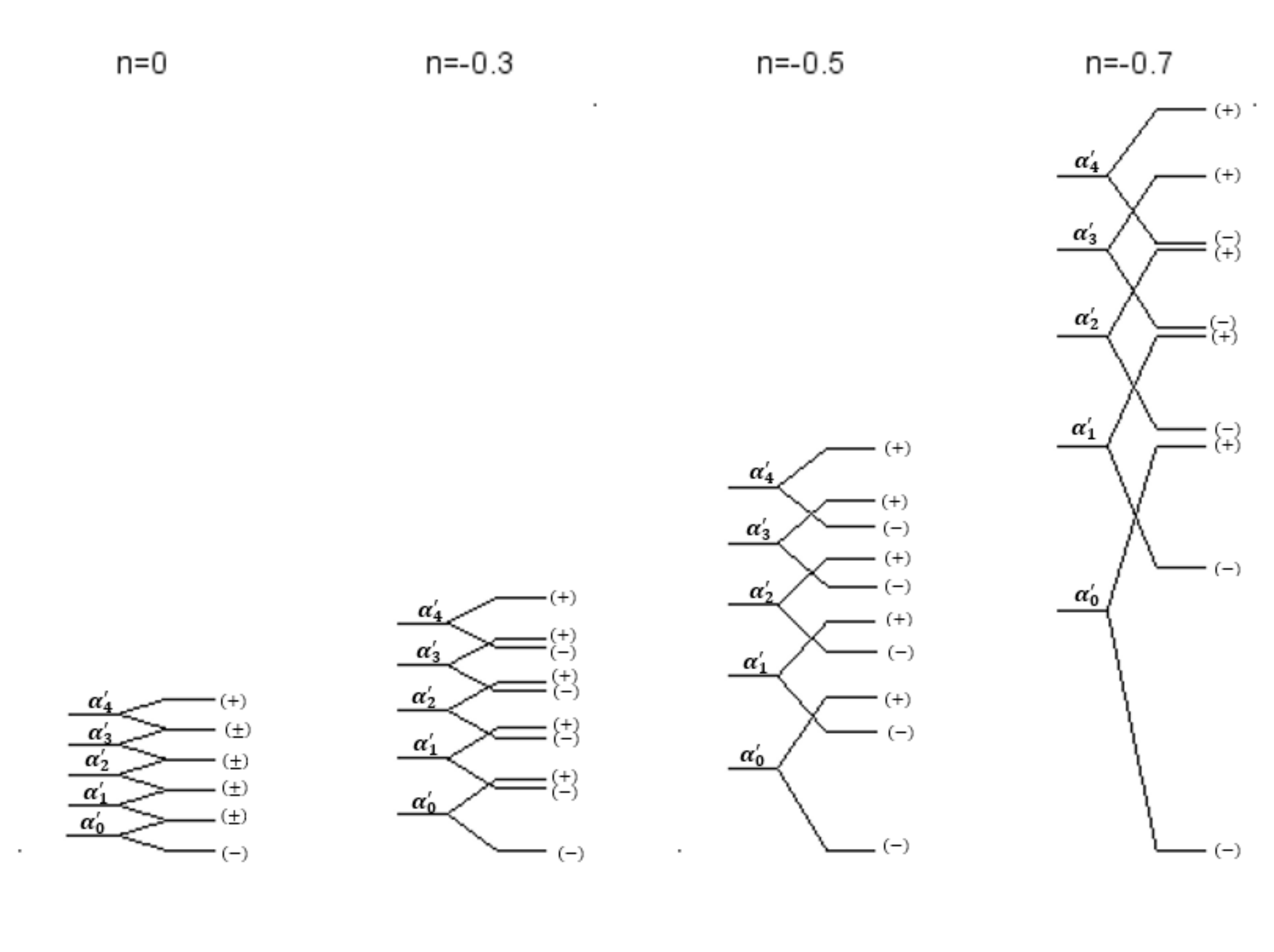}
\caption{Schematic diagram showing splitting of energy levels with constant
and varying magnetic fields for $n=0,-0.3,-0.5,-0.7$.
}
\label{schemsplit}
\end{figure*}

\begin{figure}
\centering
\includegraphics[scale=0.6]{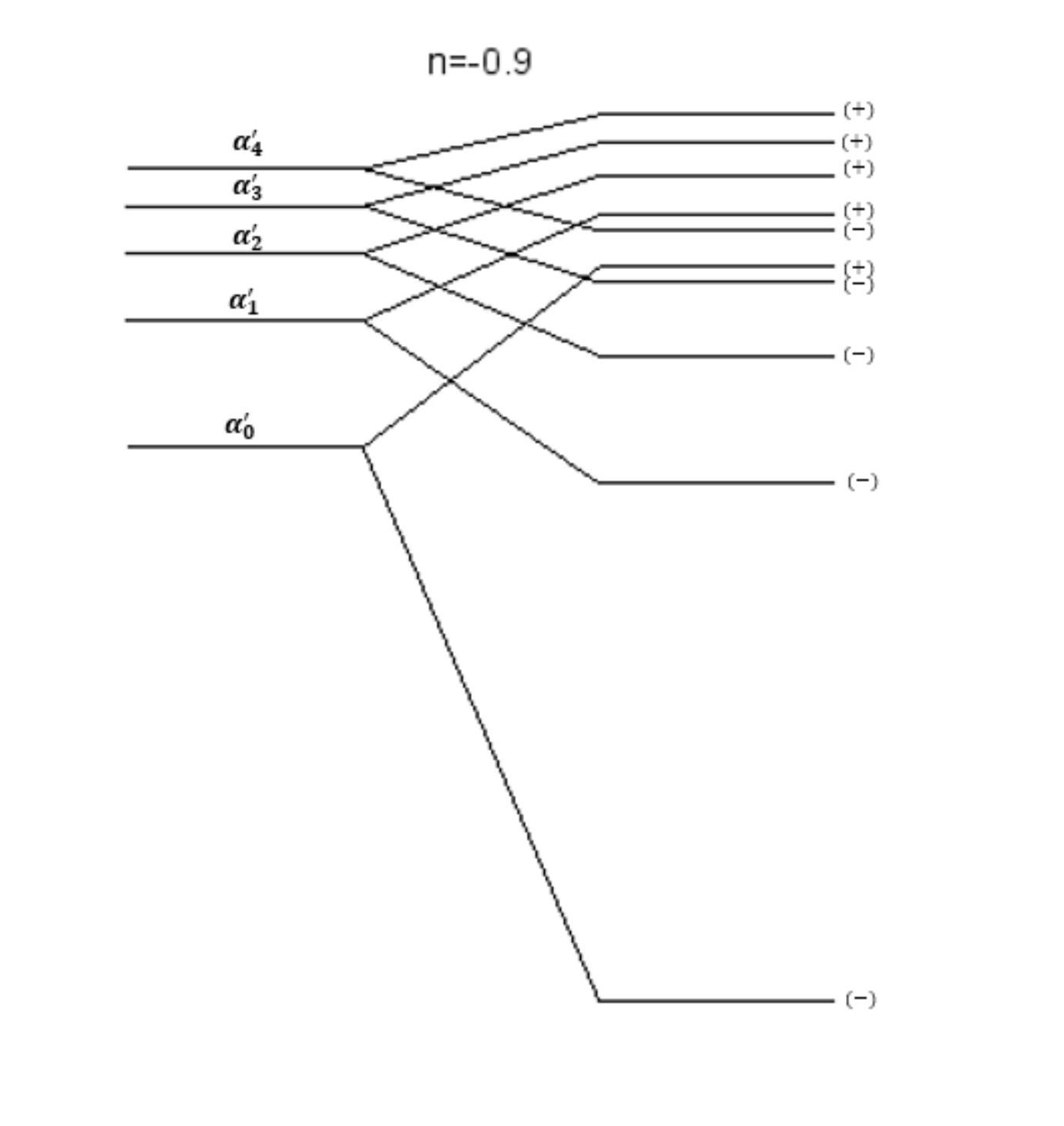}
\caption{Schematic diagram showing splitting of energy levels with $n=-0.9$.
}
\label{point9}
\end{figure}

The levels which are doubly degenerate in the presence of a constant magnetic field 
turn out to be non-degenerate when the field varies. A diagram corresponding to 
the solution of Eq. (\ref{eq10}) for the
splitting energy and lifting degeneracy with varying field as compared to the 
constant field case is shown in Figure \ref{schemsplit}. The trend of 
splitting is really a nice site for observation. The energy level corresponding 
to $+\boldsymbol{\sigma}\cdot\boldsymbol{B}$ of ground level, which 
overlaps with the energy level corresponding to $-\boldsymbol{\sigma}\cdot\boldsymbol{B}$ of first excited level for $n=0$, becomes a 
little higher than the energy level for $-\boldsymbol{\sigma}\cdot\boldsymbol{B}$ 
of first excited level for $n=-0.3$, and lies nearly in the middle of 
$-\boldsymbol{\sigma}\cdot\boldsymbol{B}$ of first and second 
excited energy levels for $n=-0.5$. This further falls in closer to the energy
level for $-\boldsymbol{\sigma}\cdot\boldsymbol{B}$ of second excited 
state for $n=-0.7$ 
In fact, for $n=-0.9$ the eigenvalue for the $+\boldsymbol{\sigma}\cdot\boldsymbol{B}$ of ground level is even larger than the 
$-\boldsymbol{\sigma}\cdot\boldsymbol{B}$ of third excited level,
as shown in Figure \ref{point9}.
 
Note that, ground level always lies at 0 
for all $n$. Thus, the physical effects arisen due to the electrons being 
in ground level only will remain unaltered if a constant field is replaced by 
a variable field or if there are little inhomogeneities within the constant 
field background.
However, other phenomena that involve with multiple Landau levels are ought 
to get modified due to unequal spacing and change of degeneracy of levels 
in non-uniform fields.  

\begin{figure}
\centering
\includegraphics[scale=0.54]{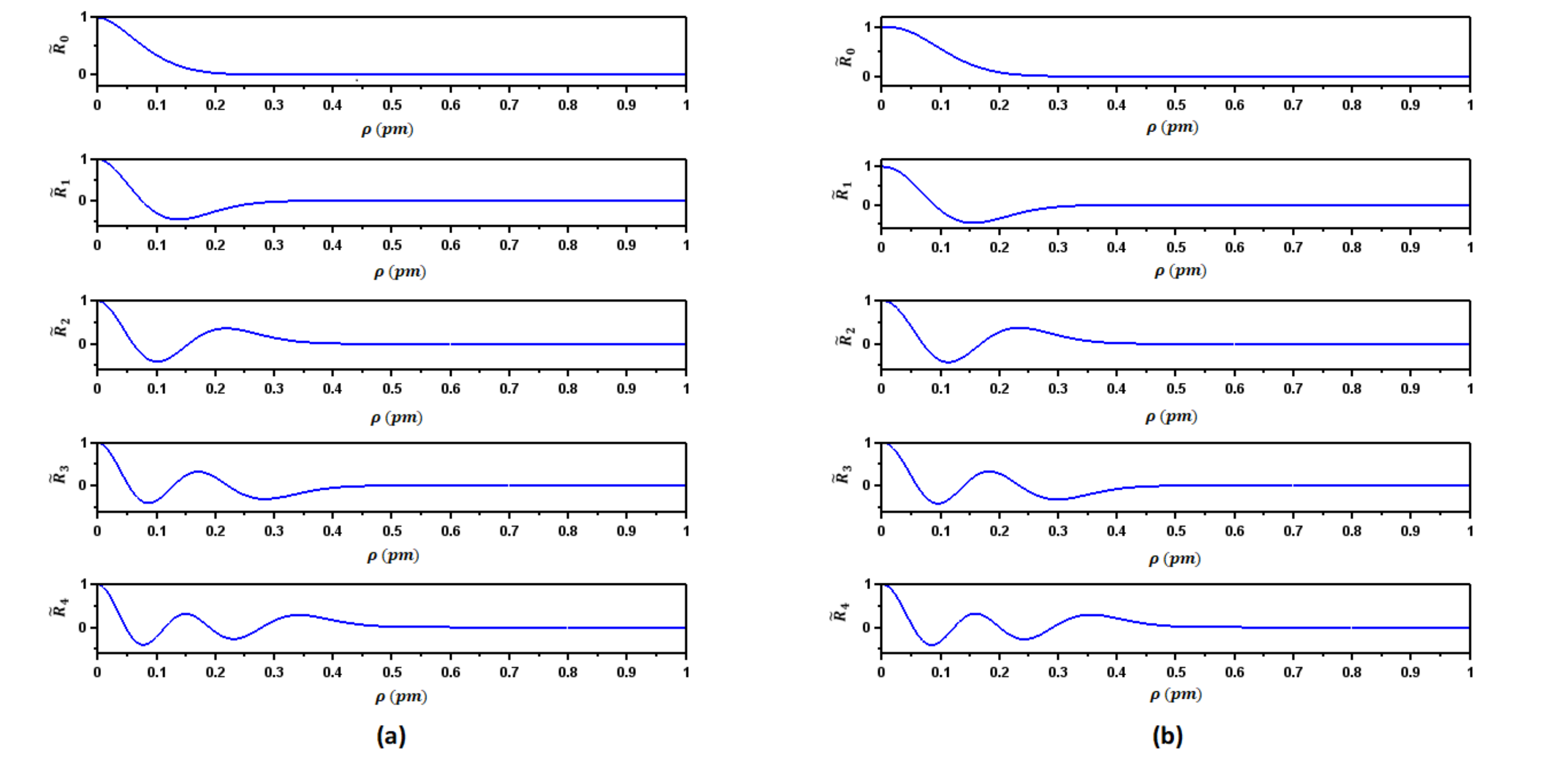}
	\caption{Wavefunctions for $B_0=10^{15}$~G pm$^{-n}$, $n=-0.3$ and $m=0$ for $\nu=0$ to $4$ 
	from the top to bottom panels respectively, for 
	(a) $-\sigma\cdot\textbf{B}$, and (b) $+\sigma\cdot\textbf{B}$.
}
\label{wavefn}
\end{figure}

Figure \ref{wavefn} shows a sample set of wavefunctions in first few levels.
It is clear that wavefunctions fully decay in the region used to determine
the eigenvalues. This ensures the correctness of eigenvalues obtained
from our computation.

\subsubsection{Dispersion relation and non-linearity}
In order to obtain the dispersion relation, we propose the ansatz for the shift of eigenvalues from the previous case, given by
\begin{equation}
\alpha_\nu = \alpha'_\nu\pm D_1~{B_0}^{D_2}~(\nu+D_3)^{D_4},
\label{eq16}
\end{equation}
where $D_1,\:D_2,\:D_3$ and $D_4$ are constants. With many trials and tribulations, we are able to obtain these constants till $n=-0.5$. However, the 
eigenvalues of very low levels (ground to third) for $+\boldsymbol{\sigma}\cdot\boldsymbol{B}$ do not 
satisfy these relations exactly, which show that the effect of 
$\pm \boldsymbol{\sigma}\cdot\boldsymbol{B}$ is not equal near the origin. 
This confirms that the effect of change in potential on the particle is 
non-linear and hence supports the power-law ansatz of our proposed dispersion 
relation. To make it lucid, there is an equal change in the potential due 
to $-\boldsymbol{\sigma}\cdot\boldsymbol{B}$ and $+\boldsymbol{\sigma}\cdot\boldsymbol{B}$, but when we compute the differences for the same with respect to 
$\alpha'_\nu$, they follow slightly different trends, which imply that the equal 
decrease and increase in potential does not have same effect proving the net 
non-linear dispersion relation for variable magnetic field.

We try to refine the constants in Eq. (\ref{eq16}) by assuming that they must 
have some particular relation with the constants of Eq. (\ref{eq14}). The relations for $n\ge -0.5$ come out to be 
\begin{equation}
\begin{matrix}
D_1 = C_3\times C_5;\\
D_2 = C_4;\\
D_3 = C_5;\\
D_4 = C_6-1.
\end{matrix}
\end{equation}

As $n$ lowers further below $-0.5$, the non-linearity in potential increases so much that 
the eigenvalues show large deviation from these relations till higher levels 
($\nu_m=10-50$). 
Hence, the net dispersion relation for $m=0$ and $n\ge-0.5$ is
\begin{equation}
\alpha_\nu = C_3\: B_0^{\frac{2}{n+2}}\:(\nu+C_5)^{\frac{2+2n}{n+2}}\left[1\pm \frac{C_5}{ (\nu+C_5)}\right].
\end{equation}

\subsubsection{$m\neq0$}

\begin{figure}
\includegraphics[scale=.7]{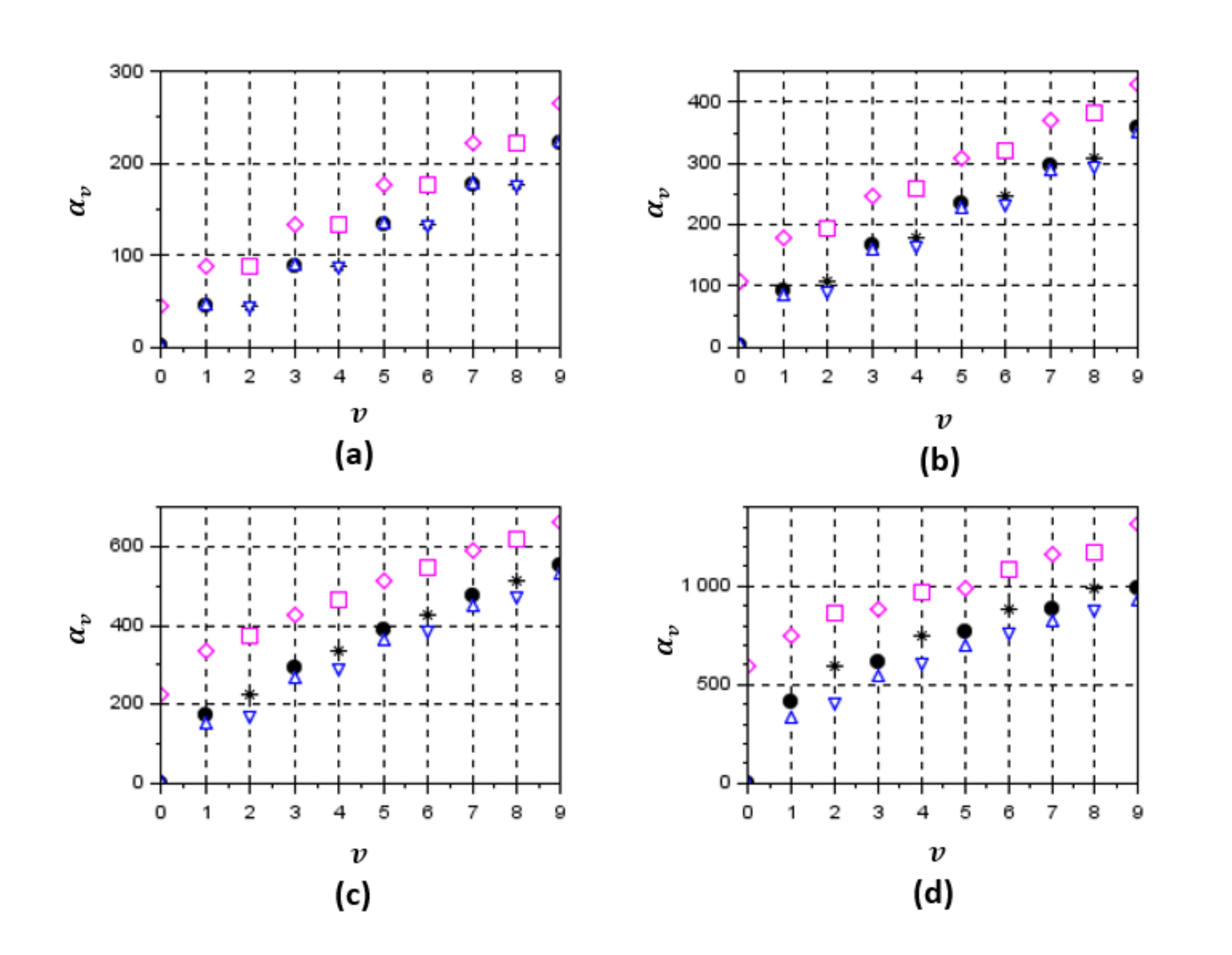}
\caption{The variation of eigenvalue with the eigen-index for $m = 0$ 
with $-\sigma.B$ (black solid circles) and +$\sigma.B$ (black asterisks);
$m=+1$ with $-\sigma.B$ (blue upward triangles) and +$\sigma.B$ (blue
downward triangles); $m=-1$ with $-\sigma.B$ (magenta diamonds) and 
 $+\sigma.B$ (magenta squares) for (a) $n=0$, (b) $n=-0.3$, (c) $n=-0.5$, and
(d) $n=-0.7$.}
\label{figwithm}
\end{figure}

When we probe the eigenvalues taking non-zero $m$, they show enthralling 
trends. We compare the eigenvalues between $m=1$ and $m=-1$ along with $m=0$ 
for $n=-0.3$, $-0.5$ and $-0.7$. As understood from Eq. (\ref{eq12}) for the case 
of constant field and discussed below Eq. (\ref{eq12}), positive $m$ does not have any impact on eigenvalues with respect to $m=0$.
Figure \ref{figwithm} shows how with decreasing $n$ (with more non-uniform field) 
eigenvalues for $m=0$ and
$m=1$ along with $m=-1$ become distinct at a given eigen-index. 
It is clearly seen that as $n$ decreases, the difference between the eigenvalues for 
$m=0$ and $m=1$ increases.

Above behavior may be inferred to be an effect stemming from the rotation of 
particles and field behavior. Generally, rotation in the direction to the magnetic field is easier
and to the opposite direction difficult. The case with of $m\neq0$ implies that the 
electron has some angular momentum, which further implies its rotational motion. 
A positive $m$ means rotation in the direction to the magnetic field and negative 
implies opposite. When the magnetic field is homogeneous, a rotation in the direction 
to the magnetic field is not expected to require any extra force when there is no change 
of field magnitude (no force due to magnetic pressure), hence no change in 
energy. However, to rotate a particle opposite to its natural direction, 
extra force is required, hence raising of energy. For an inhomogeneous field, 
as the magnitude of field changes (here decreases with distance), the particle 
has to overcome the force due to magnetic pressure, even if the direction is same as 
of the magnetic field. Hence, its own energy dissipates, leading to the less energy 
to align with the magnetic field.

\section{\label{eos} Modification to thermodynamic properties and equation of state of cold degenerate electron gas}
The main impact of the variation of LQ in the presence of varying magnetic 
field established above is to the systems where field varies drastically over the spatial 
scale.
Since there is a huge difference between central and surface magnetic fields 
in the astrophysical bodies like white dwarfs and neutron stars, 
their realistic properties should be determined in the presence of variable magnetic 
fields, in place of an approximate constant field. However, the quantum effect is
important, and the application of preceding discussion works out in stellar astrophysics,
only if the variation of strong fields is in the length scale of gyromagnetic
radius. Otherwise, even if LQ might play an important role to determine the underlying EoS and 
stellar structure depending on the field strength, uniform magnetic field based results suffice.
In high densities, matter in white dwarfs and 
neutron stars turns out to be degenerate and their EoSs 
play indispensable role to determine the underlying stellar properties. In the 
presence of strong magnetic field, such a highly dense matter may get 
influenced by LQ, depending on the field strength, composition and density. 
The variation of magnetic field can be chosen appropriately by considering 
suitable $n$ in our chosen field ansatz, once central and surface magnetic fields are known or at best
anticipated.

As emphasized above, modified LQ, based on non-uniform magnetic field,
is useful in determining its effect in EoS only in pm scale.
The main effect of magnetic field via LQ is to modify the available density of states 
for electrons. For a constant field, the difference in energy levels is 
constant, but for a variable magnetic field case, the energy difference 
between levels no longer remains constant, as shown in Figure \ref{schemsplit}. 
Also there are separate sets of energy levels for spin-up and spin-down 
electrons. 
In the presence of variable magnetic field 
EoS can be found out as follows.

Considering only one kind of electrons at a time, say spin-down, the number of 
states per unit volume in a momentum interval $\Delta p_z$  for a Landau level $\nu$ 
for non-uniform energy levels is given by (generalized from 
\cite{PhysRevD.86.042001})
\begin{equation}
\frac{\pi }{h^3}\left(\tilde{P}_{\nu +1}-\tilde{P}_{\nu}\right)\Delta p_z.
\label{eq31}
\end{equation}
For a constant magnetic field and all electrons, the above expression is 
amended with a degeneracy factor $g_{\nu}$ of Landau levels, where $g_{\nu}=1$ 
for the ground state and $g_{\nu}=2$ for other states. However the situation
is different for a non-uniform field.

Let us define $\left(\tilde{P}_{\nu +1}-\tilde{P}_{\nu}\right)_{\pm} = D(\nu)_{\pm}$. Therefore, the electron density of states in the absence of magnetic field $\frac{2}{h^3}\int d^3p$ is replaced by
\begin{equation}
\frac{2\pi}{h^3}D(\nu)_{\pm}\int dp_z
\label{eq321}
\end{equation}
in the case of a non-zero magnetic field.

In order to calculate the electron number density $n_e$ at zero temperature, we have to evaluate the integral in Eq. (\ref{eq321}) from $p_z = 0$ to $p_F(\nu)$, which is the Fermi momentum of the Landau level $\nu$, and obtain
\begin{equation}
	{n_e}_{\pm}= \sum_{\nu =0}^{\nu _m}\frac{2\pi}{h^3} D(\nu)_\pm p_F(\nu).
\label{eq32}
\end{equation}
The Fermi energy $E_F$ of the electrons for the Landau level $\nu$ is given by
\begin{equation}
E_F^2= m_e^2c^4+p_F(\nu)^2c^2+\tilde{P}(\nu)c^2.
\label{eq33}
\end{equation}
The upper limit $\nu_m$ of the summation, corresponding to the upper limit
of levels, in Eq. (\ref{eq32}) is derived from 
the condition that $p_F^2(\nu)\geq0$, which implies
\begin{equation}
\tilde{P}(\nu)c^2\leq E_F^2-m_e^2c^4
\label{eq34}
\end{equation}
or
\begin{equation}
\alpha_{\nu_m} = \epsilon_{F\:max}^2 - 1.
\label{eq341}
\end{equation}
Hence, total electron density taking into account both the spins of electron is
\begin{align}
	\nonumber
	n_e &= {n_e}_+ + {n_e}_- \\ \nonumber 
	&= \frac{1}{(2\pi)^2\lambda_e^3}\left(\sum_{\nu=0}^{\nu=\nu_{m-}}\beta _-(\nu)x_{F-}(\nu)+\sum_{\nu=0}^{\nu=\nu_{m+}}\beta _+(\nu)x_{F+}(\nu)\right),\\
\label{eq35}
\end{align}
where
$+$ sign indicates spin-up and $-$ sign spin-down, $x_{F}=p_F/m_e c$,
\begin{equation}
x_{F\pm}(\nu) = \left[\epsilon_F^2 - (1+\alpha_{\pm}(\nu)\right]^{\frac{1}{2}}
\label{eq36}
\end{equation}
and
\begin{equation}
\beta_{\pm} = \left(\alpha_{\pm}(\nu+1) - \alpha_{\pm}(\nu-1)\right)/2.
\end{equation}
The electron energy density at zero temperature is

\begin{eqnarray}
	\nonumber
	&&\varepsilon_e= \frac{1}{(2\pi)^2\lambda_e^3}\left(\sum_{\nu=0}^{\nu=\nu_{m-}}\beta_-(\nu)\int_0^{x_{F-}(\nu)}E_{\nu,p_z}dx_z\right. +\left. \sum_{\nu=0}^{\nu=\nu_{m+}}\beta_+(\nu)\int_0^{x_{F+}(\nu)}E_{\nu,p_z}dx_z\right)\\ \nonumber
	&&\qquad= \frac{m_ec^2}{(2\pi)^2\lambda_e^3}\left(\sum_{\nu=0}^{\nu=\nu_{m-}}\beta_-(\nu)(1+\alpha_-(\nu))f_1\left[\frac{x_{F-}(\nu)}{(1+\alpha_-(\nu))^{1/2}}\right]\right. \\  && \qquad \qquad \qquad+\left.
\sum_{\nu=0}^{\nu=\nu_{m+}}\beta_+(\nu)(1+\alpha_+(\nu))f_1\left[\frac{x_{F+}(\nu)}{(1+\alpha_+(\nu))^{1/2}}\right]\right),
\label{eq38}
\end{eqnarray}

where
\begin{equation}
f_1(z) = \frac{1}{2}\left(z\sqrt{1+z^2} + ln (z+\sqrt{1+z^2})\right)
\end{equation}
and $E_{\nu,p_z}$ is the quantized energy levels defined in, e.g., Eq. (\ref{eqP}).
The pressure of an electron gas is given by
\begin{eqnarray}
	\nonumber
	&&P_e= n_e^2\frac{d}{dn_e}\left(\frac{\varepsilon_e}{n_e}\right) = -\varepsilon_e+n_eE_F\\ \nonumber
	&&\qquad=\frac{m_ec^2}{(2\pi)^2\lambda_e^3}\left(\sum_{\nu=0}^{\nu=\nu_{m-}}\beta_-(\nu)(1+\alpha_-(\nu))f_2\left[\frac{x_{F-}(\nu)}{(1+\alpha_-(\nu))^{1/2}}\right] \right. \\ && \qquad \qquad \qquad+\left.
\sum_{\nu=0}^{\nu=\nu_{m+}}\beta_+(\nu)(1+\alpha_+(\nu))f_2\left[\frac{x_{F+}(\nu)}{(1+\alpha_+(\nu))^{1/2}}\right]\right),
\label{eq39}
\end{eqnarray}
where
\begin{equation}
f_2(z) = \frac{1}{2}\left(z\sqrt{1+z^2} - ln (z+\sqrt{1+z^2})\right).
\end{equation}

\begin{figure*}
\centering
	\includegraphics[scale=.45]{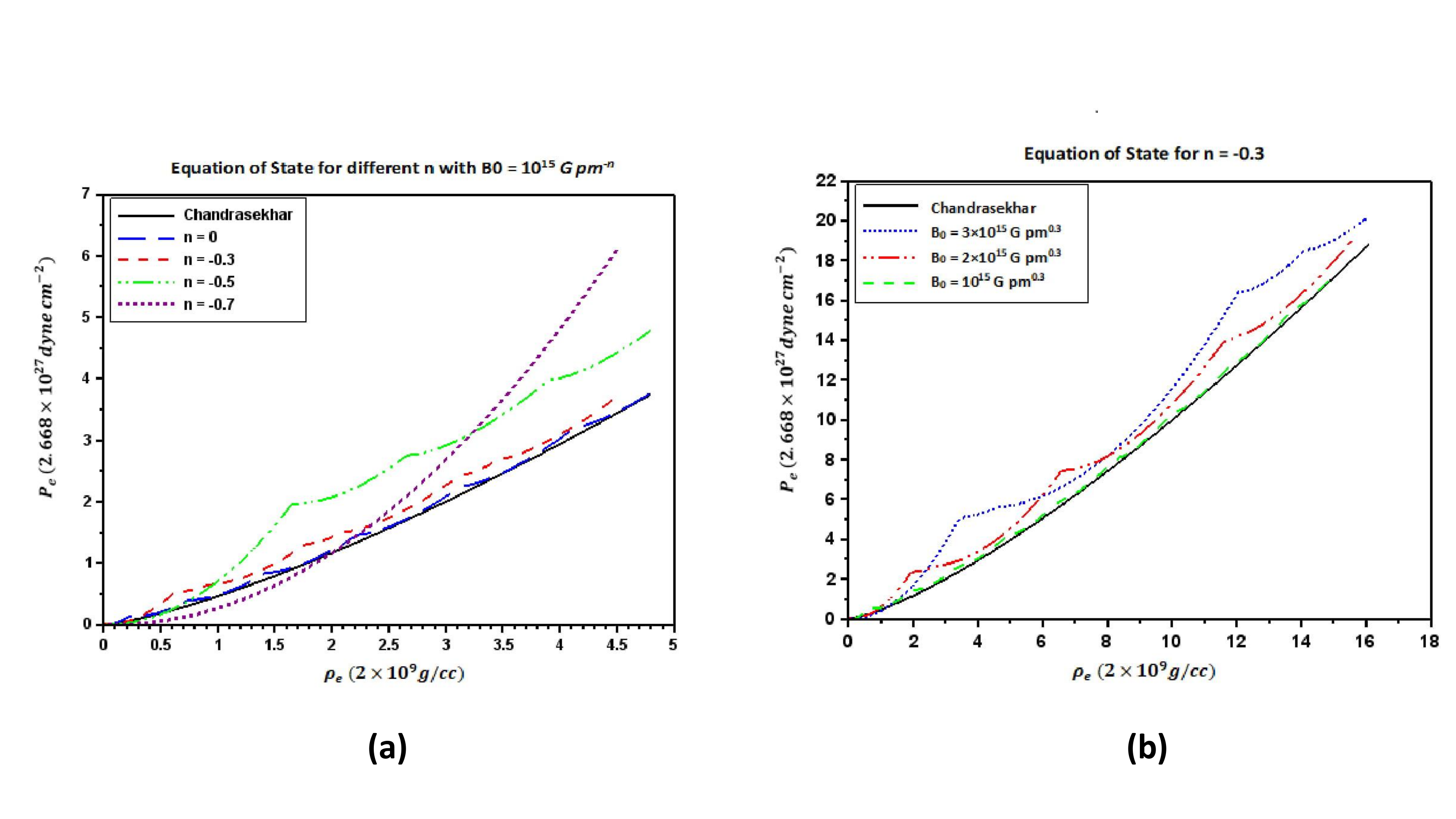}
	\caption{(a) Equation of state for $n=0$ with $\epsilon_{Fmax}=17$ (long-dashed blue line),
	$n=-0.3$ with $\epsilon_{Fmax}=17$ (dashed red line), $n=-0.5$ with $\epsilon_{Fmax}=18$ 
	(dot-dashed green line) and $n=-0.7$ with $\epsilon_{Fmax}=15$ (dotted violate line), for $B_0=10^{15}$ G pm$^{-n}$, 
	along with Chandrasekhar's result with $\epsilon_{Fmax}=17$ (solid black line), (b) Equation of state for $B_0=10^{15}$~G pm$^{-n}$ (dashed green line), 
	$2\times 10^{15}$~G pm$^{-n}$ (dot-dashed
	red line) and $3\times 10^{15}$~G pm$^{-n}$ (dotted blue line) along with Chandrasekhar's result (solid black line)
	for $n=-0.3$ with $\epsilon_{Fmax}=25$, when pressure is in units of $2.668\times 10^{27}$ erg cm$^{-3}$ and mass density in units of $2\times 10^9$ gm cm$^{-3}$.
}
\label{eosn}
\end{figure*}

We know that with the change in allowed number of levels in a system for a given $\epsilon_{Fmax}$, EoS changes significantly. As the number of level increases, the pressure decreases and increases for a given density, which are respectively called softer and harder/stiffer EoS, at a high and low densities respectively (see \cite{PhysRevD.86.042001} for the example of constant field). Figure \ref{eosn}(a) shows EoS for various $n$. With the decrease in $n$, EoS becomes stiffer at a high density and softer at a low density, indicating lesser number of allowed levels in the system for low $n$. Figure \ref{eosn}(b) shows how EoS becomes stiffer, at the high density regime,
with increasing $B_0$ for a fixed $n$. This is as per the expectation as stronger field leads to
the more LQ effect with less number of levels populated, deviating the results more from the
nonmagnetic case.

Note importantly that above EoSs shown in Figure \ref{eosn} are applicable to a B-WD till the radius
from the center where field does not decay significantly and hence LQ is still valid. On the other hand,
depending on $n$, field decays with the radial coordinate in the pm scale and hence a given EoS does not
remain valid for a B-WD beyond a scale of the order of pm.

\section{\label{appl} Implications}
\subsection{\label{app11} Mass--Radius Relation of Magnetized White Dwarfs}
An immediate astrophysical implication of \textbf{LQ} is to the mass--radius
relation of (highly) magnetized white dwarfs. As mentioned in the Introduction
(e.g., \cite{2013PhRvL.110g1102D,2020MNRAS.496..894G}), strong magnetic
field can significantly modify the mass-radius relation due to LQ 
as well as classical Lorentz force effects. 
However, it has been argued sometime \cite{dipankar} that LQ effect is not important in 
controlling stellar structure of white dwarfs and only Lorentz force would
suffice the same. Here we plan to check if \textbf{LQ} has any impact on the white dwarf stellar structure.

For the present purpose, we consider a sample field profile in cylindrical
polar coordinates as
\begin{eqnarray}
	\textbf{B}&=&B_0\hat{z},\,\,\,{\rm for}\,\,\,\rho < 850~{\rm km},\\ \nonumber
	\textbf{B}&=&B_0\left(\frac{\rho}{\rm 1~km}\right)^{-0.37}\hat{z},\,\,\,\,{\rm for}\,\,\,\rho < 900~{\rm km},\\ \nonumber
	\textbf{B}&=&B_0\left(\frac{\rho}{\rm 1~km}\right)^{-0.99}\hat{z},\,\,\,\,\,{\rm otherwise},\\ \nonumber
	\label{bprof2}
\end{eqnarray}
so that $(\textbf{B}\cdot\nabla)\textbf{B}=0$ and $\nabla\cdot\textbf{B}=0$.
Therefore, the nonrotating white dwarfs will be spherical in shape. 
Hence, in spherical polar coordinates with $\theta=\pi/2$, the field 
profile is given by
\begin{eqnarray}
	\textbf{B}&=&-B_0\hat{\theta},\,\,\,{\rm for}\,\,\,r < 850~{\rm km},\\ \nonumber
	\textbf{B}&=&-B_0\left(\frac{r}{\rm 1~km}\right)^{-0.37}\hat{\theta},\,\,\,\,{\rm for}\,\,\,r < 900~{\rm km},\\ \nonumber
	\textbf{B}&=&-B_0\left(\frac{r}{\rm 1~km}\right)^{-0.99}\hat{\theta},\,\,\,\,\,\,{\rm otherwise}.\\ \nonumber
	\label{bprof}
\end{eqnarray}
This profile assures (based on the solution for the stellar structure given below) that at the surface the 
field is restricted to be around $10^{12}$ G when $B_0=2\times 10^{15}$ G.

Therefore, the mass and radius of a white dwarf can be obtained by solving 
\begin{equation}
	\frac{d}{dr}\left(P_e+\frac{B^2}{8\pi}\right)=-\frac{GM(r)
	(\rho_e+\rho_B)}{r^2},
	\label{msbal}
\end{equation}
\begin{equation}
	\frac{dM(r)}{dr}=4\pi r^2(\rho_e+\rho_B),
\end{equation}
where $\rho_B$ is the magnetic density, $B^2=\textbf{B}\cdot\textbf{B}$, $\rho_e=n_e m_p\mu_e$, $m_p$ is the 
mass of proton, $\mu_e$ is the mean molecular weight per electron and
$G$ is Newton's gravitation constant. Here for $r<850$~km, EoS would be Landau quantized for 
$B_0=2\times 10^{15}$ G, but 
for a uniform magnetic field. For $r\ge 850$~km, the field decays to a lower strength so that Chandrasekhar's
nonmagnetic EoS suffices. Only at the interface around 850~km, non-uniform field based LQ applies in EoS,
but in a very tiny region. Hence, for the present example, LQ based on uniform field practically influences EoS.

\begin{figure}
\centering
	\includegraphics[scale=.4]{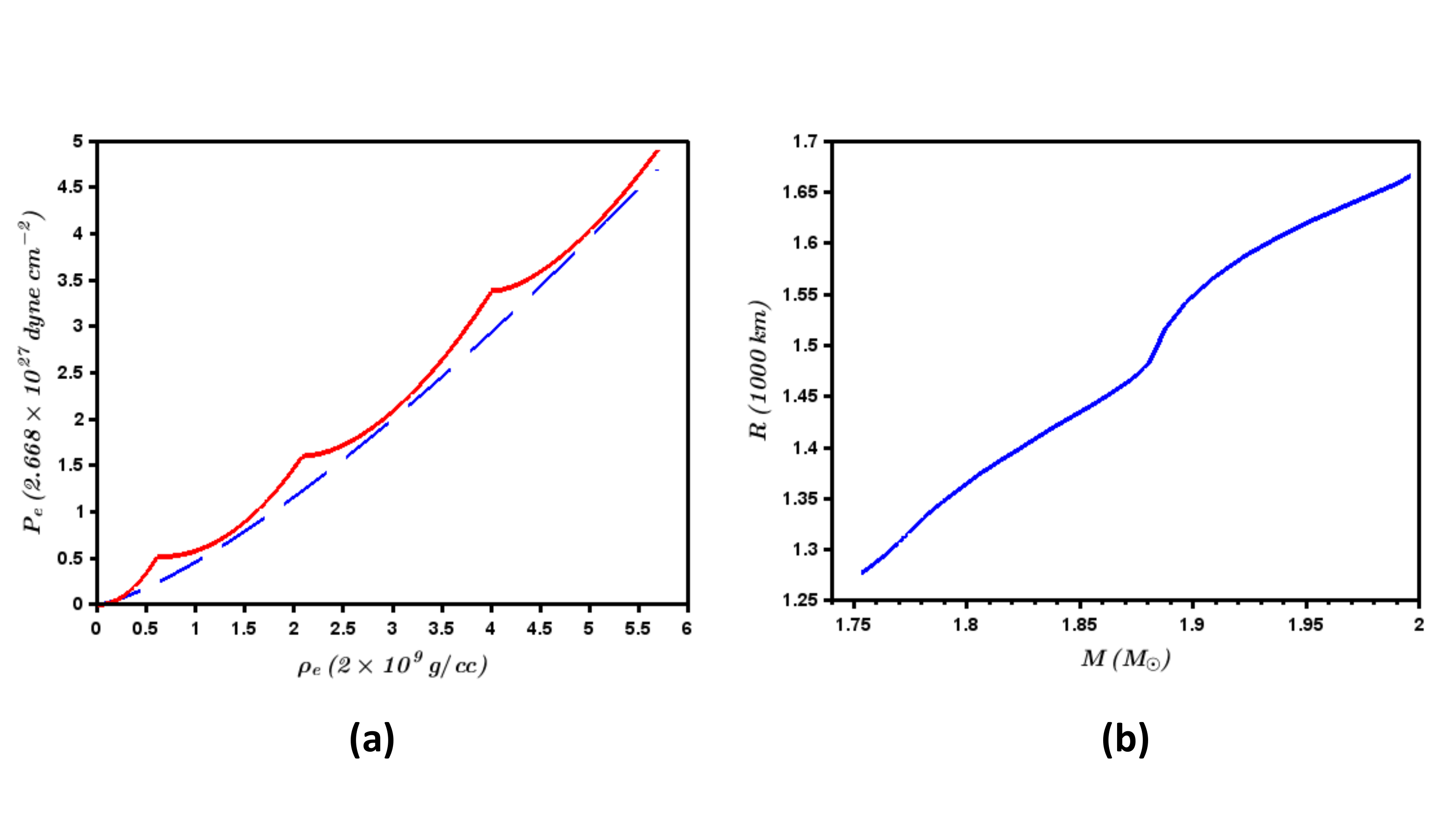}
	\caption{(a) EoS for constant magnetic field ($n=0$)
	of $B_{cent}=2\times 10^{15}$ G (red solid line) describing the central
	region of white dwarfs (see Eq.
        \ref{bprof2}) and Chandrasekhar's EoS (dashed blue line) 
	describing otherwise for $\epsilon_{Fmax}=18$, (b) corresponding mass--radius relation.
}
\label{Eos}
\end{figure}

\begin{figure}
\centering
	\includegraphics[scale=.7]{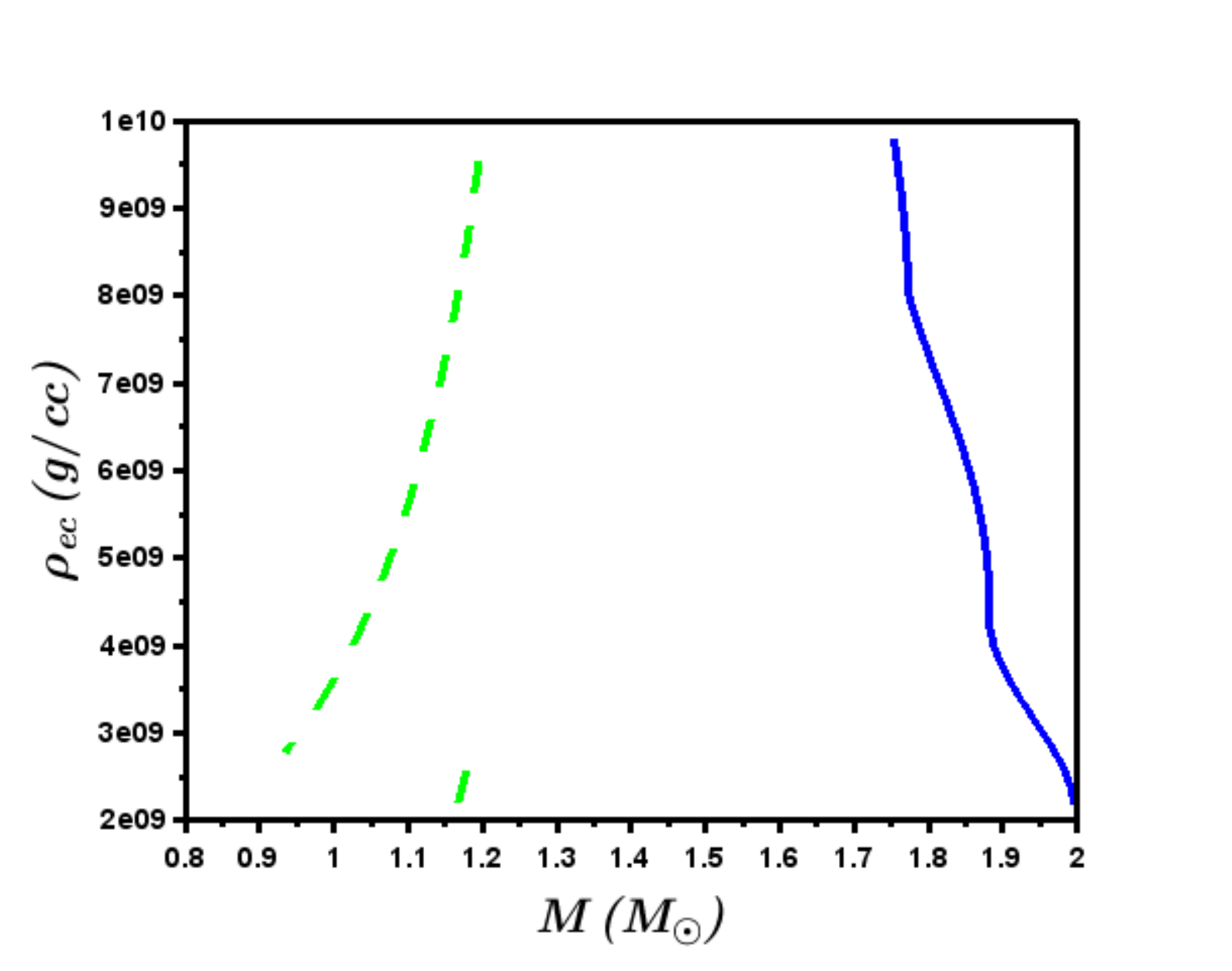}
	\caption{Comparison of $M-\rho_{ec}$ relation (solid blue line) with
	that of the hypothetical case with Chandrasekhar's EoS but Lorentz
	force intact (dashed green line).
}
\label{mdenboth}
\end{figure}

Figure \ref{Eos}(a) shows
EoS for a constant magnetic field describing B-WD from centre to $r<850$~km,
along with Chandrasekhar's nonmagnetic EoS which is applicable in $r\ge 850$~km
for the star with profile given by Eq. (\ref{bprof2}).
Figures \ref{Eos}(b)
and \ref{mdenboth} show that 
the mass turns out to be significantly super-Chandrasekhar for the present field
profile and the mass-limit may arise from the upper limit of density, e.g., 
arisen due to pycnonuclear reactions, neutron drip etc. We plan to explore this in detail
in a future work, particularly the deviation from the mass-radius trend of 
Chandrasekhar with increasing $B_0$. Note that in principle with decreasing 
$\rho_c$, maximum field in a B-WD, i.e. $B_0$, should be decreasing. However,
for uniformity and the convenience of comparison and computation, we have
kept $B_0$ same for all the stars in the sample computations. This shows
apparent decreasing mass with increasing $\rho_c$.

In Figure \ref{mdenboth} we assess how important the LQ effect over the Lorentz
force is, at least for the chosen profile. We find that a hypothetical case
with Lorentz force but without LQ, i.e. with Chandrasekhar's EoS, leads 
the mass to restrict below the Chandrasekhar-limit. This is understood as the
high density B-WDs have smaller radius and according to the chosen profile
given by Eq. (\ref{bprof2}) field remains constant throughout or almost throughout,
hence there is no (significant) Lorentz force. Therefore, the mass is restricted 
mostly based on Chandrasekhar's EoS. However, lower density B-WDs, 
having radius larger than 850~km with the field varying in the outer region,
exhibit Lorentz force, which is however not adequate enough to increase mass.
Hence, LQ plays a significant role to bring in super-Chandrasekhar white dwarfs.
Note that the hypothetical mass--radius relation exhibits two discrete 
branches (one being around $1.15-1.2M_\odot$). This break of continuity in
the mass is due to the sharp change in field due to its transition from
constant to varying trends, which however does not arise in the realistic 
case with LQ effect included, as shown in Figure \ref{Eos}(b).

Of course, the chosen field profile is just a test sample, in particular to
facilitate the exploration of LQ, keeping other physics intact. 
This however establishes that for a realistic
case, the LQ effect should not be neglected at high magnetic fields.

\subsection{\label{app12} Quantum Speed Limit}

Quantum speed of particle determines how fast it transits from one energy level to another. It has direct influence on the processing speed of quantum information. It was shown by Villamizar and Duzzoini \cite{PhysRevA.92.042106} that for an electron in ground state with the up spin, the maximum quantum speed, also known as \textit{quantum speed limit}, irrespective of the magnitude of magnetic field is $0.2407c$, if the magnetic field is uniform. We apply the same idea in 
the regime of non-uniform magnetic field.

The wavefunction of an electron with the up spin in state $\nu$ is given by
\begin{equation}
\Psi = e^{\frac{-iE_{\nu}t}{\hbar}}\psi_{\nu},
\end{equation}
such that
\begin{equation}
\psi_{\nu} = e^{i\left(m\phi+\frac{p_{z}}{\hbar}z\right)}\begin{bmatrix}
R_{\nu +}(\rho)\\
-R_{\nu +}(\rho)\\
\end{bmatrix}.
\end{equation}
Let the initial state of the spin-up electron be the superposition of two consecutive states, ground and first excited states, with $m=0$ and $p_z=0$, given by
\begin{equation}
 \Psi(\rho,0) = \frac{1}{\sqrt{2}}\left[\psi_0(\rho)+\psi_1(\rho)\right],
 \end{equation} 
 where the respective energies are $E_0$ and $E_1$.
 
 Consider the evolution of wavefunction from ground state to first excited state. The minimum time of evolution is given by the Mandelstam-Tamm (MT) bound 
 \cite{mandelstam1945energy}
 \begin{equation}
 T_{min} = \frac{\pi\hbar}{2\Delta H},
 \end{equation}
 where
 \begin{equation}
 \Delta H = \frac{E_1 - E_0}{2}
 \end{equation}
 in this case.
 
 The radial displacement of particle in time $T_{min}$ is 
 \begin{align}
\rho_{disp} &= |\langle\rho\rangle_{T_{min}}-\langle\rho\rangle_0| \\ 
&=2\left\vert\int^\infty_0 \rho D_S(\rho)d\rho\: \right\vert,
\label{eq5q}
\end{align}
where 
\begin{equation}
D_s(\rho) = \psi^{+}_{0}\:\rho\:\psi_{1}.
\label{eq6q}
\end{equation}
Thus, quantum speed of electron is given by 
\begin{equation}
\tilde{\rm v} = \frac{\rho_{disp}}{T_{min}}.
\end{equation}

In order to determine quantum speed limit, we choose large $B_0$ ($=10^{16}~{\rm G\:pm}^{-n}$) such that on changing $B_0$ further, there is not much change in 
quantum speed for a given $n$. 
\begin{figure}
\centering
\includegraphics[scale=.7]{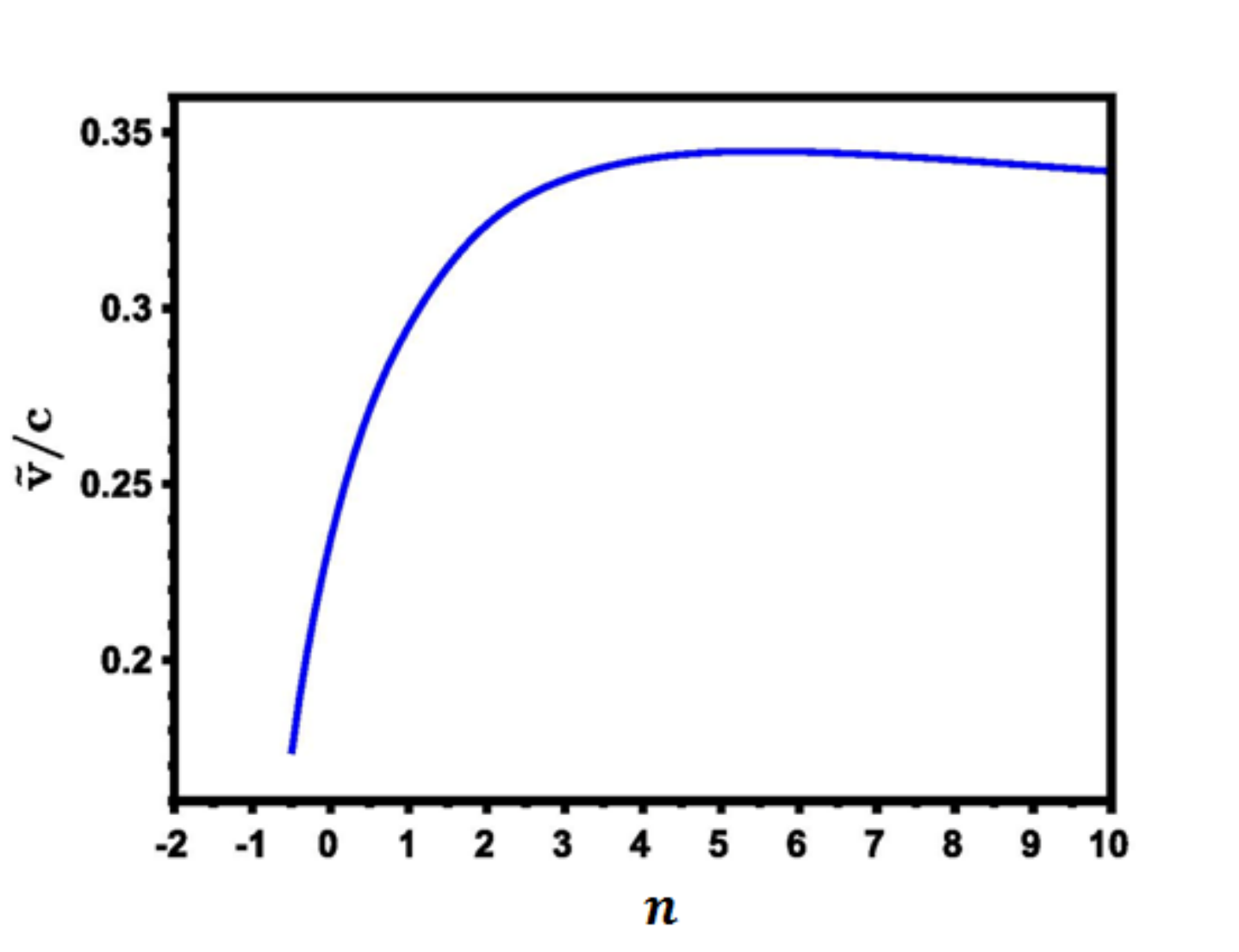}
\caption{Variation of quantum speed of spin-up electron for transition from 
	ground state to first excited state with different $n$ at $B_0 = 10^{16}~{\rm G\:pm}^{-n}$.}
\label{qsl}
\end{figure}

\begin{figure}
\centering
	\includegraphics[scale=.6]{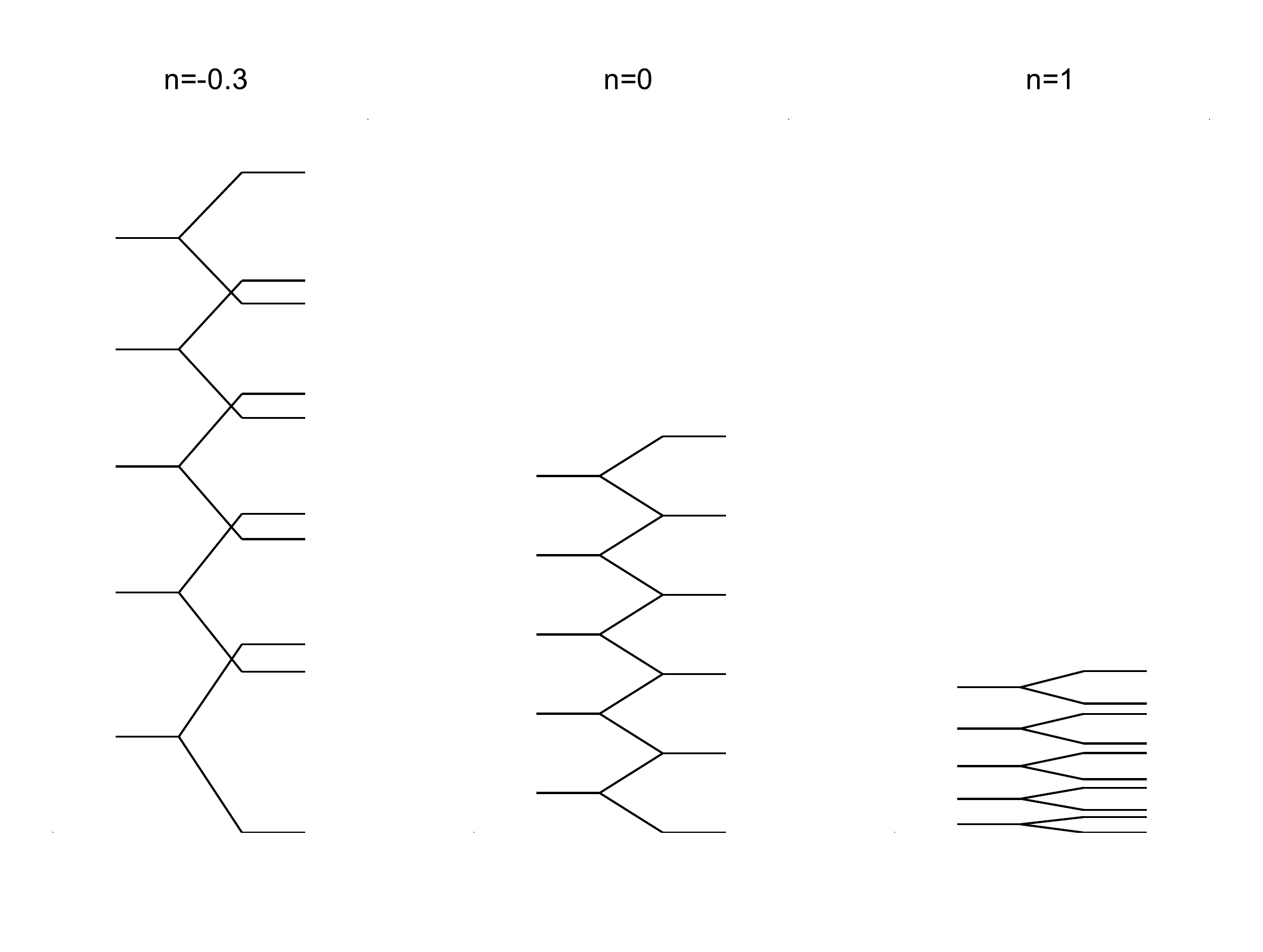}
\caption{Comparison of schematic representation of the energy level 
	splitting between the magnetic field with $n<0$ and that with
	$n>0$, along with a case of uniform magnetic field ($n=0$).}
\label{npm}
\end{figure}

As it can be seen from Figure \ref{qsl}, quantum speed of electron increases with increasing $n$, reaches maximum and then begins to decrease. The quantum 
speed limit increases compared to its value in a uniform magnetic field
($n=0$) for $n>0$. This is related to different rearrangements of 
energy levels lifting degeneracy between the fields with $n<0$ and $n>0$, shown in 
Figure \ref{npm}. Thus, if we could trap an electron in a magnetic field which is spatially increasing in 
magnitude even linearly ($n=1$) within a small scale, then we can achieve 
a higher speed of transition of electron. This can be extremely useful in 
faster processing of quantum information in the presence of variable magnetic 
field as compared to the uniform field. We plan to investigate this
application of variable magnetic field in detail in a future work.

\section{\label{sec7} Conclusion}

The LQ and the underlying dispersion relation with varying magnetic fields is
a unique problem on its own. When the field is constant, the problem is 
nothing but a harmonic oscillator, whose analytical solution for 
energy is well-known. However for a varying field, the situation is quite 
different and difficult. Unless the magnetic field or more precisely the 
underlying vector potential follows a specific
profile, e.g. a power-law variation, even a semi-analytical solution seems
to be very difficult. We have chosen a simple power-law variation of the 
magnetic field and the corresponding vector potential, so that its satisfies
no magnetic monopole condition and also magnetic tension vanishes. The latter
helps applying this result in stellar physics easily.

For the ease of comparing with the constant field case, we develop the underlying 
quantum mechanics in the cylindrical coordinate system, which however can 
easily be recast for spherical coordinates. We have obtained a very important
result. Due to the variation of magnetic field, the degeneracy in energy levels, 
as known for the constant field, is lifted and there is unique alignment of levels of spin-up and spin-down electrons depending on the nature of change of magnetic field. The result is not difficult to
understand. As the field magnitude changes at each point, the LQ effect keeps varying
at each point as well, which leads to non-overlapping energy levels as they are for
constant field, hence lifting the degeneracy. 

The non-uniform magnetic field has a wide range of applications in 
the both decaying as well as in rising regimes.
If we consider the decaying magnetic field profile,
the above result importantly has a significant consequence to the EoS of the 
magnetized degenerate electron gas. For a similar Fermi energy,
at the low density regime, pressure decreases compared to the constant field case,
while it is opposite in the high density regime, for a given density. 

Modification to the EoS due to LQ further leads to the increase 
of white dwarf mass 
significantly compared to the mass without field. While an effect of constant
field on to the white dwarf mass arises solely due to the LQ, the varying field
also add to the Lorentz force. 
However, depending on the nature of field variation,
the LQ effect plays an indispensable role to determine the stellar 
structure and the mass of white dwarfs. Our sample
field profile chosen for a test case is not far from reality.
Thus, it establishes that in a realistic situation, the LQ in a white 
dwarf with a strong magnetic field is an important effect.
On the other hand, the spatially growing magnetic field could be 
proven useful in quantum information where we can achieve higher 
quantum speed of particle in the presence of variable magnetic field.

In general, LQ effects in variable magnetic field 
can be applied to a variety of physical systems ranging from astrophysics
to quantum information to high energy physics to condensed matter. Suitable modifications to the 
effects such as quantum Hall effect in the 
presence of non-uniform magnetic field could give rise to unexplored but 
interesting experimental consequences.

\section*{Acknowledgments}
The authors thank Subhashish Banerjee of IIT Jodhpur and Diptiman Sen of IISc
for reading the preliminary draft, discussion and suggestions. Also thanks
are due to Surajit Kalita of IISc for cross-checking some of the mass--radius
relations and discussion, and Debabrata Deb of IISc for useful comments on 
central magnetic field. S.A. thanks her undergraduate teacher S. N. Sandhya 
of Miranda House, University of Delhi, for teaching the coding techniques 
used in this work.
\section*{Author Contributions}
B.M. proposed the problem and supervised the project; S.A. set up the numerical model
and solved it; B.M. and S.A. formulated the problem, analyzed the numerical results and wrote the manuscript. G.G. verified the results and helped in improving the manuscript by adding physical subtleties to the work.

\begin{appendix}
\section{Obtaining Dirac and Maxwell's equations from Lagrangian}

The total Lagrangian density of the system of electrons of wave-function $\psi$ in the presence of electro-magnetic field is
\begin{equation}
        {\cal L} = \bar{\psi}\left(i\gamma^{\mu}D_{\mu}-m \right)\psi -\frac{1}{16\pi} F_{\mu\nu}F^{\mu\nu},
\end{equation}
where
\begin{equation}
        D_{\mu} = \partial _{\mu} - ieA_{\mu},~~~F_{\mu\nu}=\partial_\mu A_\nu-\partial_\nu A_\mu,
\end{equation}
in the units $\hbar=c=1$, where $\mu$ runs from 0 to 3.
Using Lagrangian equations of motion
\begin{equation}
        \partial_\nu\left(\frac{\partial\cal{L}}{\partial(\partial_{\nu}{\bar \psi})}\right) - \frac{\partial\cal{L}}{\partial{\bar \psi}}=0,
\end{equation}
we obtain
\begin{equation}
\left( i\gamma^{\mu}D_{\mu}-m\right)\psi=0,
	\label{dir}
\end{equation}
which is the Dirac equation. Further using
\begin{equation}
        \partial_\nu\left(\frac{\partial\cal{L}}{\partial(\partial_{\nu}A_{\mu})} \right)- \frac{\partial\cal{L}}{\partial A_{\mu}}=0,
\end{equation}
we obtain
\begin{equation}
        \partial_{\nu}F^{\nu\mu}-4\pi j^{\mu}=0,
\label{maxj}
\end{equation}
which is the inhomogeneous Maxwell's equation,
where $j^{\mu}$ is the current density, given by
\begin{equation}
j^{\mu} = e\bar{\psi}\gamma^{\mu}\psi.
\end{equation}
For the time-independent magnetic field with vanishing electric field,
Eq. (\ref{maxj}) reduces to 
\begin{equation}
	\nabla\times{\bf B}=4\pi{\bf J}.
\label{maxj2}
\end{equation}
For the present purpose of Landau quantization in the presence of varying magnetic fields,
we have solved the Eq. (\ref{dir}) above and obtained eigenvalues.
On the other hand, for stellar structure, we have considered Eq. (\ref{maxj2}) in order to 
introduce Lorentz force proportional to ${\bf J}\times{\bf B}$.
\end{appendix}





 \bibliography{newbib}

\nolinenumbers

\end{document}